# Spatially resolved fluorescence of caesium lead halide perovskite supercrystals reveals quasi-atomic behavior of nanocrystals

Dmitry Lapkin[1,*], Christopher Kirsch[2,*], Jonas Hiller[2,*], Denis Andrienko[3], Dameli Assalauova[1], Kai Braun[2], Jerome Carnis[1], Young Yong Kim[1], Mukunda Mandal[3], Andre Maier[2,4], Alfred J. Meixner[2,4], Nastasia Mukharamova[1], Marcus Scheele[2,4,+], Frank Schreiber[4,5], Michael Sprung[1], Jan Wahl[2], Sophia Westendorf[2], Ivan A. Zaluzhnyy[5], Ivan A. Vartanyants[1,6,+]

1. *Deutsches Elektronen-Synchrotron DESY, Notkestraße 85, 22607 Hamburg, Germany*

2. *Institut für Physikalische und Theoretische Chemie, Universität Tübingen, Auf der Morgenstelle 18, 72076 Tübingen, Germany*

3. *Max Planck Institute for Polymer Research, Ackermannweg 10, 55128 Mainz, Germany*

4. *Center for Light-Matter Interaction, Sensors & Analytics LISA+, Universität Tübingen, Auf der Morgenstelle 15, D-72076 Tübingen, Germany*

5. *Institut für Angewandte Physik, Universität Tübingen, Auf der Morgenstelle 10, 72076 Tübingen, Germany*

6. *National Research Nuclear University MEPhI (Moscow Engineering Physics Institute), Kashirskoe shosse 31, 115409 Moscow, Russia*

\* These authors contributed equally

+To whom correspondence should be addressed

**Abstract**

We correlate spatially resolved fluorescence (-lifetime) measurements with X-ray nanodiffraction to reveal surface defects in supercrystals of self-assembled caesium lead halide perovskite nanocrystals and study their effect on the fluorescence properties. Upon comparison with density functional modelling, we show that a loss in structural coherence, an increasing atomic misalignment between adjacent nanocrystals, and growing compressive strain near the surface of the supercrystal are responsible for the observed fluorescence blueshift and decreased fluorescence lifetimes. Such surface defect-related optical properties extend the frequently assumed analogy between atoms and nanocrystals as so-called quasi-atoms. Our results emphasize the importance of minimizing strain during the self-assembly of perovskite nanocrystals into supercrystals for lighting application such as superfluorescent emitters.

**Introduction**

Advances in the self-assembly of colloidal nanocrystals (NCs) from solution into three-dimensional arrays with long-range order have enabled the design of microscopic "supercrystals" that approach the structural precision of atomic single crystals.[1] The individual NCs, which are the building blocks of a supercrystal, are often regarded as "artificial atoms", and hence analogies between atomic crystals and such supercrystals have been made.[2,3] NC supercrystals are susceptible to doping,[4] and they can exhibit exceptional mechanical properties,[5] quasicrystal formation,[2] enhanced electronic coupling,[6] and engineered phonon modes.[7] In view of the recent progress in exploiting the massive structural coherence in NC supercrystals to generate collective optoelectronic properties,[8–10] a critical question remains whether this artificial atom analogy can be extended towards the optical properties of NC supercrystals. Due to surface dangling bonds and surface reconstruction, even the purest and most carefully prepared atomic crystals are not structurally perfect.[11,12] For *atomic* crystals, such surface defects strongly affect the fluorescence spectra, lifetime and quantum yield.[13–17] For supercrystals, this is much less understood.

In this work, we show that in close analogy to atomic crystals, $CsPbBr_2Cl$ and $CsPbBr_3$ NC supercrystals exhibit structural distortions near their surfaces which significantly alter their fluorescence properties. This finding is of high relevance for the application of these materials as tunable, bright emitters with superfluorescent behavior.[8–10] Superfluorescence is a key property for the design of spectrally ultra-pure laser sources[18] or highly efficient light-harvesting systems.[19] Recent quantum chemical simulations have suggested that structural disorder in $CsPbBr_3$ supercrystals and its effect on the thermal decoherence plays a pivotal role in the efficiency of the superfluorescence.[20] Previous structural investigations of ensembles of $CsPbBr_3$ supercrystals by grazing-incidence small angle X-ray scattering (SAXS) indicated a primitive unit cell with slight tetragonal distortion,[21] and wide-angle X-ray scattering (WAXS) showed a high degree of structural coherence.[22] Electron microscopy of individual supercrystals revealed a frequent occurrence of local defects in the supercrystals, such as isolated NC vacancies.[23] Confocal fluorescence microscopy of individual $CsPbBr_3$ supercrystals displayed spatial variations in the fluorescence peak wavelength and intensity, indicating that local structural inhomogeneities may substantially affect the fluorescence properties of the entire supercrystal.[24] Our approach is based on simultaneous WAXS and SAXS measurements with a nano-focused beam to probe the structural defects and crystallographic orientation of the supercrystal and the constituting NCs on a local level with dimensions of ~3 µm and 7 – 9 nm, respectively.[25–27] By correlation with diffraction-limited confocal fluorescence microscopy and modelling with density functional theory (DFT) we present proof that compressive strain, a loss of structural coherence and an increasing atomic misalignment between adjacent nanocrystals at the edges of $CsPbBr_2Cl$ NC supercrystals are responsible for a blueshifted emission and decrease of the fluorescence lifetimes.

## Results

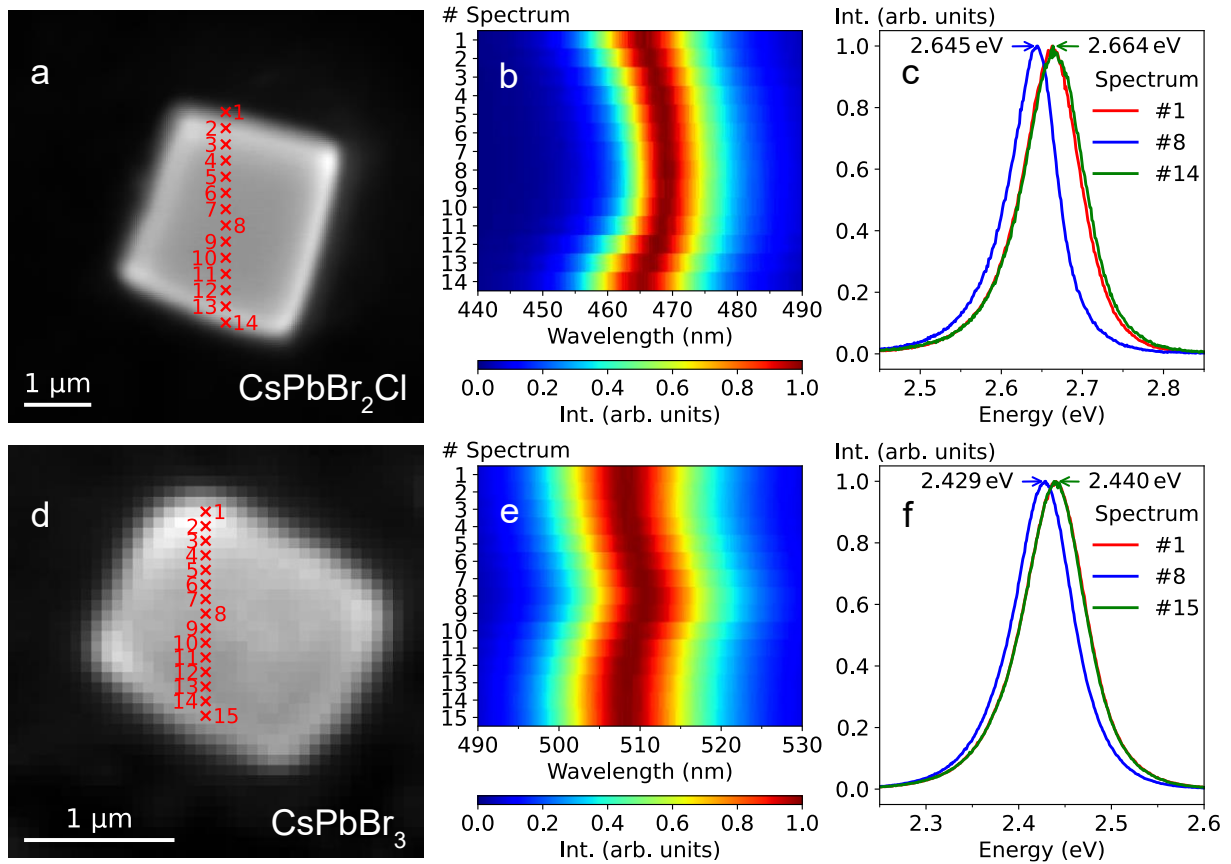

**Figure 1: Spatially resolved fluorescence.** (a) Optical micrograph of a CsPbBr$_2$Cl NC supercrystal. Positions of the measured photoluminescence spectra are indicated. (b) The corresponding normalized spectra. (c) Selected normalized spectra, acquired at the edges and the center of the supercrystal. (d-f) Corresponding data for a CsPbBr$_3$ supercrystal.

We study self-assembled CsPbBr$_2$Cl and CsPbBr$_3$ NC supercrystals on glass substrates (see Methods for details on synthesis and self-assembly of NCs). Spatially resolved photoluminescence spectra of the NC supercrystals under 405 nm excitation in a confocal laser scanning microscope with a step size of 250 nm and 100 nm, respectively, are shown in **Figure 1**. When approaching an edge of the supercrystal, we find a continuous blueshift of the emission peak wavelength. This blueshift is strongest for relatively small (few µm edge length) and highly faceted supercrystals, where it reaches up to 20 meV for CsPbBr$_2$Cl. We observe the same blue-shifting behavior for supercrystals composed of CsPbBr$_3$ NCs, although to a lesser extent (up to 12 meV).

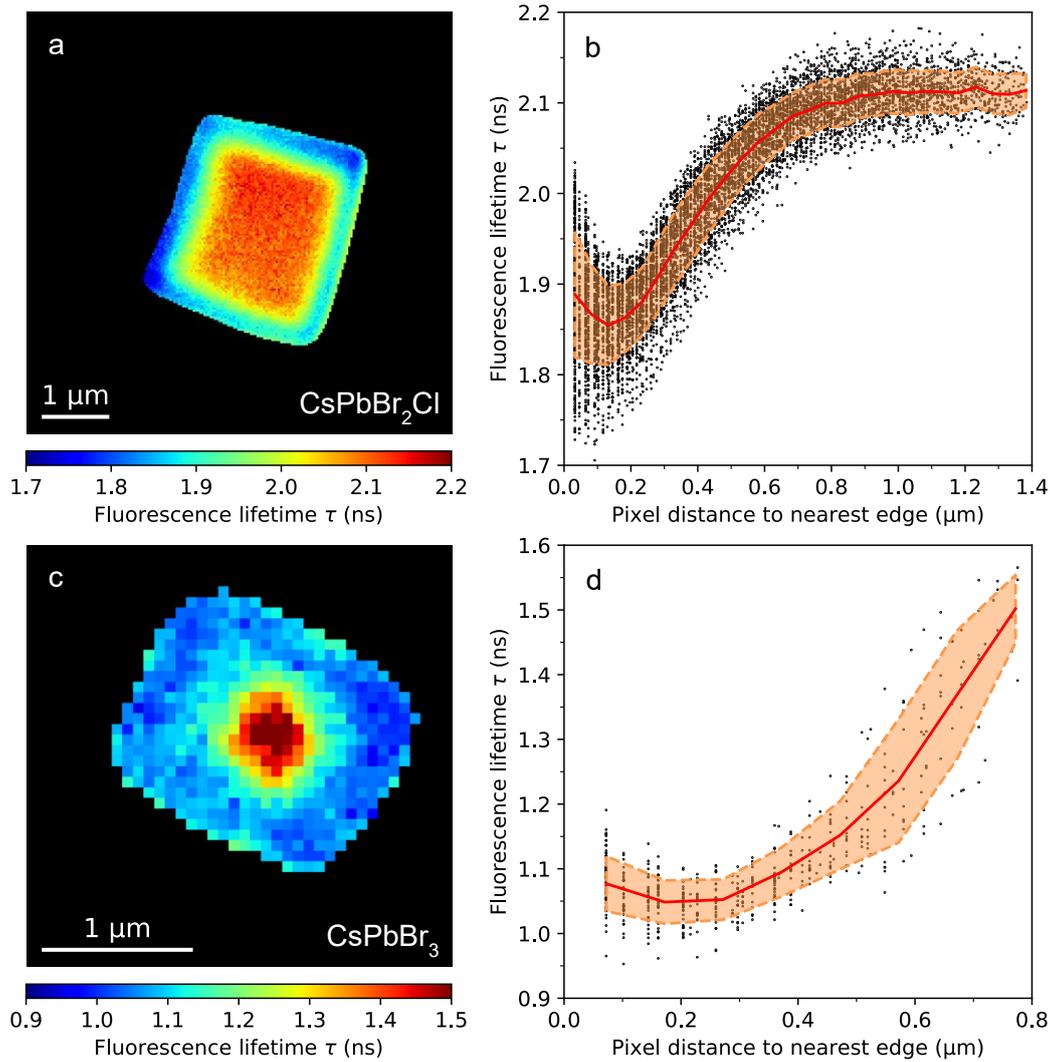

**Figure 2: Spatially resolved fluorescence lifetime imaging. (a)** Fluorescence lifetime image of a CsPbBr$_2$Cl NC supercrystal obtained by fitting the experimental time-resolved fluorescence with a monoexponential decay function. **(b)** Fluorescence lifetime values obtained at each pixel inside the supercrystal as a function of the distance to the nearest edge, where the red line shows the mean value, and the dashed lines indicate the confidence interval of ±σ. **(c-d)** Analogous results for a CsPbBr$_3$ NC supercrystal.

In **Figure 2**, we display fluorescence lifetime images of self-assembled CsPbBr$_2$Cl and CsPbBr$_3$ supercrystals measured on glass substrates with a lateral resolution of 200 nm under 405 nm excitation. For both supercrystal compositions, we obtain good fits of the experimental time-resolved fluorescence by pixel-by-pixel monoexponential reconvolution using an instrument response function acquired on a clean glass coverslip (Supplementary **Figure S2, S3**). In the case of supercrystals composed of $CsPbBr_2Cl$ NCs, we measure typical fluorescence lifetimes around 2.2 ns in the center which decrease by approximately 20% when scanning from the center of a supercrystal towards its edges. Supercrystals composed of $CsPbBr_3$ NCs exhibit typical lifetime values around 1.5 ns in the center, which shorten by approximately 30% when approaching the edges. We note that this holds true only for freshly prepared NC supercrystals. After several days of exposure to air, the trend in the spatially resolved τ-values is reversed in that such aged supercrystals exhibit longer lifetimes at the edges. However, the overall blueshift of the fluorescence peak wavelength towards the edges is preserved.

To correlate the fluorescence data with the structure of the supercrystals, we carry out X-ray synchrotron measurements by SAXS and WAXS at PETRA III facility (Hamburg, Germany) (see **Figure 3a** and Methods for details). Using a $400 \times 400$ nm$^2$ X-ray beam, we perform a spatially resolved scan of a typical $CsPbBr_2Cl$ NC supercrystal on a Kapton substrate. While the results presented here are for one typical supercrystal, examples of more supercrystals are provided in the Supporting information (**Section S7**). First, all individual patterns are integrated to obtain the average structure. The averaged background-corrected WAXS and SAXS diffraction patterns are shown in **Fig. 3b** and **3c**, correspondingly. The signal in the WAXS region contains three orders of Bragg peaks from the atomic lattice (**Fig. 3b**), and the SAXS region (shown enlarged in **Fig. 3c**) displays several orders of Bragg peak from the supercrystal. A real-space map of the scan based on the integrated SAXS intensity at $q < 2$ nm$^{-1}$ is shown in **Fig. 3d**. The map represents a square area of high intensity corresponding to a single supercrystal. For comparison, we display a scanning electron micrograph of a similar supercrystal (see inset in **Figure 3a** and **Figure S6**) from which we determine an average NC diameter of 7.3±0.4 nm and an interparticle distance of 2.5±0.5 nm. For strongly faceted supercrystals, the NC diameter is rather uniform over the whole crystal. For less faceted supercrystals, occasional ensembles of smaller NCs are found in the vicinity of the edges. However, the spatial extent of such smaller NC populations is always limited to ~200 nm (see Supporting Information, **Section S3**).

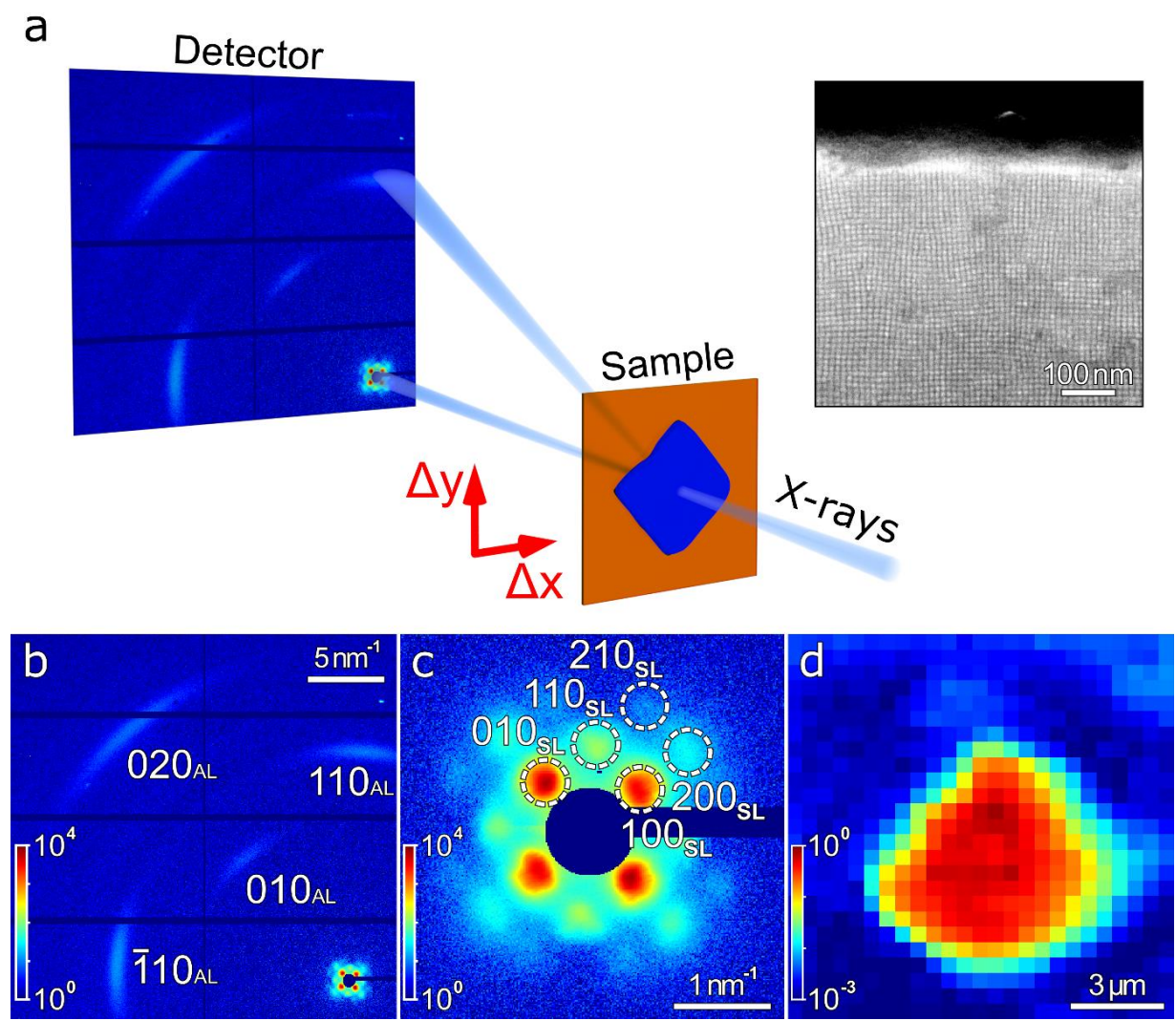

**Figure 3: Spatially resolved X-ray nanodiffraction experiment and average diffraction patterns.** (**a**) Scheme of the X-ray experiment. EIGER X 4M 2D detector is positioned downstream from the sample. The arrows show the directions Δx and Δy of spatial scanning. Inset (top right): a SEM micrograph of the $CsPbBr_2Cl$ NC supercrystal. (**b**) Average diffraction pattern for a supercrystal. Several orders of WAXS and SAXS Bragg peaks from the atomic

and supercrystal structure, respectively, are well visible. The WAXS Bragg peaks are indexed using pseudo-cubic notation. (**c**) Enlarged SAXS region of the averaged diffraction pattern. The Bragg peaks are indexed according to a simple cubic structure. (**d**) Diffraction map for a scan based on the integrated intensity of the SAXS diffraction patterns at $q < 2$ nm$^{-1}$. The pixel size (the step size) is 500 nm.

The average diffraction pattern in the WAXS region (see **Fig. 3b**) contains four prominent Bragg peaks, originating from the atomic lattice (AL) of the NCs. Their radial positions at $q = 10.93$ nm$^{-1}$, 15.44 nm$^{-1}$, and 21.90 nm$^{-1}$ (see Supporting Information, **Figure S8**) can be attributed to a cubic AL. We note that although a cubic phase for CsPbBr$_2$Cl has been reported,[28,29] the most stable phase at room temperature is expected to be orthorhombic. Due to the small NC size and the resulting broadening of the Bragg peaks, it is impossible to distinguish between these two very similar structures. Thus, we use a pseudocubic notation to index the WAXS peaks: 110 and 002 orthorhombic peaks correspond to 100$_{AL}$ pseudocubic peak, 112 and 200 – to 110$_{AL}$, and 220, 004 – to 200$_{AL}$ peaks. The present peaks and their azimuthal positions indicate a primary orientation of the NCs along the [001]$_{AL}$ axis with respect to the incident beam. We find the unit cell parameter to be $a_{AL} = 0.575 \pm 0.003$ nm, which is in good agreement with previously reported values for CsPbBr$_2$Cl.[30] From the peak broadening, we extract the NC size (*d*) and lattice distortion (*g*) using the Williamson-Hall method with $d = 6.8 \pm 0.1$ nm and $g = 2.3 \pm 0.1\%$ (see Supporting Information, **Section S4**). The obtained NC size is in good agreement with the SEM results.

The SAXS pattern in **Figure 3c** represents the typical 4-fold pattern of a simple cubic lattice oriented along the [001]$_{SC}$ axis with four visible orders of Bragg peaks that can be attributed to 100$_{SC}$, 110$_{SC}$, 200$_{SC}$, and 210$_{SC}$ reflections of the supercrystal of NCs. We determine an average unit cell parameter of $a_{SC} = 9.9 \pm 0.4$ nm. Considering the NC size obtained by SEM, we obtain an interparticle distance of 2.6±0.4 nm, which is in a good agreement with the SEM result (2.5±0.5 nm). All crystallographic axes of the NCs are aligned with the corresponding axes of the supercrystal (e.g. $[100]_{AL}\|[100]_{SC}$ and $[010]_{AL}\|[010]_{SC}$), which is consistent with Ref. 23.

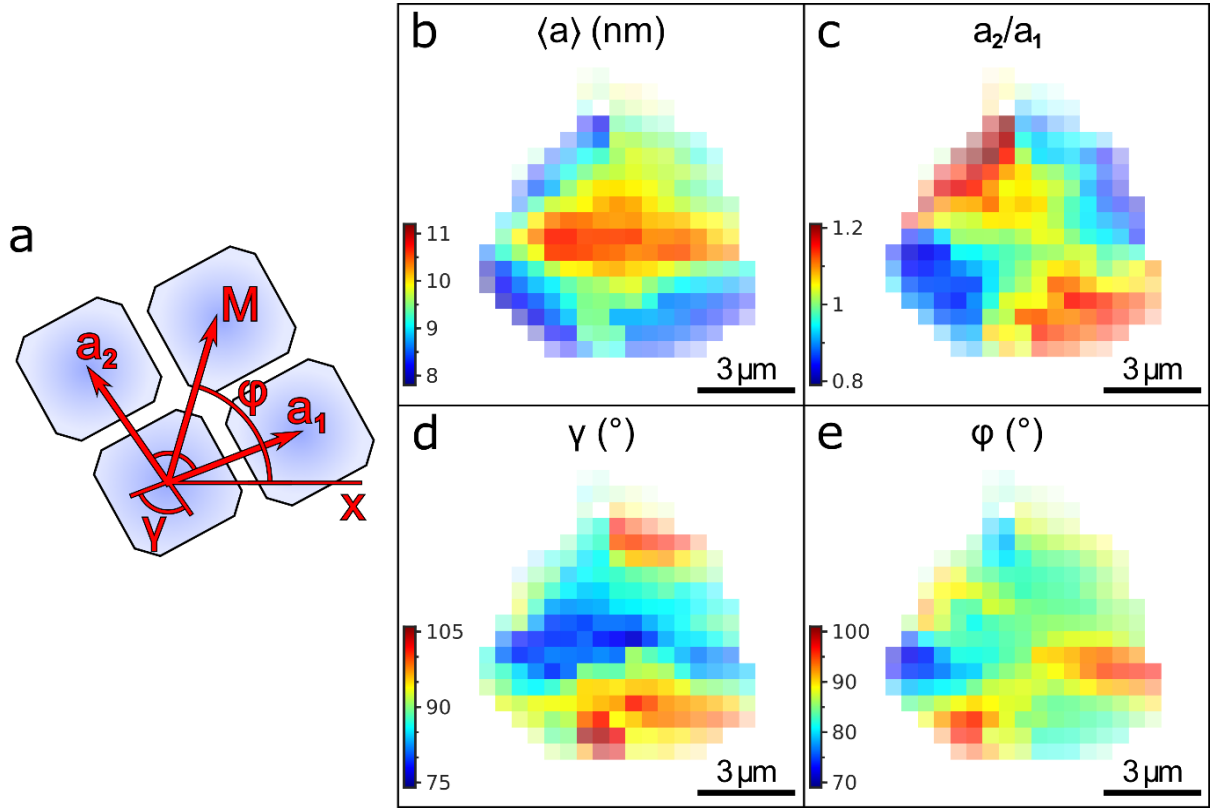

**Figure 4: Spatially resolved SAXS.** (**a**) Definition of the geometrical parameters of a superlattice unit cell: the basis vectors $\boldsymbol{a_1}$ and $\boldsymbol{a_2}$ with the angle $\gamma$ between them, and the mean line $M$ between the basis vectors at the angle $\varphi$; (**b**) average unit cell parameter $\langle a \rangle = (a_1+a_2)/2$; (**c**) ratio $a_2/a_1$ of the unit cell parameters along the basis vectors $\boldsymbol{a_2}$ and $\boldsymbol{a_1}$; (**d**) angle $\gamma$ between the basis vectors $\boldsymbol{a_1}$ and $\boldsymbol{a_2}$; (**e**) azimuthal position $\varphi$ of the mean line $M$ between the basis vectors $\boldsymbol{a_1}$ and $\boldsymbol{a_2}$. The pixel size in (b-e) is 500 nm.

Analyzing individual SAXS patterns from different locations on the supercrystal, we find substantial local deviations from the average structure (see Supporting Information, **Figure S10** for examples of single diffraction patterns). To illustrate this, from the Bragg peak positions, we extract the basis vectors $a_1$ and $a_2$, the angle $\gamma$ between them, and the average azimuthal position $\varphi$, which are defined in **Figure 4a** (see the Methods section for details). As depicted in **Fig. 4b**, the mean unit cell parameter is largest in the center of the supercrystal with 10.7 nm and smallest at the edges with 7.8 nm. Although both unit cell parameters $a_1$ and $a_2$ decrease at the edges (see Supporting Information, **Figure S14**, for separate maps of $a_1$ and $a_2$ values), we observe that this lattice contraction is anisotropic. The ratio of the in-plane unit cell parameters $a_2/a_1$ differs from unity by ±20% in such a way that the NC spacing in the directions along the nearest supercrystal boundary is smaller than normal to it, as shown in **Fig. 4c**. We note that the mean value $\langle a \rangle = 9.4 \pm 0.7$ nm is slightly smaller than the unit cell parameters extracted from the average diffraction pattern. We attribute this to the low intensity of scattering from the supercrystal edges, which reduces their contribution to the average pattern. We do not observe a clear trend in the size of the SAXS Bragg peaks (see Supporting Information, **Figure S13**, for the maps). The instrumental peak broadening, determined by the incident X-ray beam size is about 0.015 nm$^{-1}$ (FWHM). The observed peak sizes are much larger and vary in the range from 0.05 nm$^{-1}$ to 0.2 nm$^{-1}$ and, as such, they depend mainly on the superlattice distortion. The characteristic lengthscale on which this distortion evolves is, most probably, smaller than the incident beam. Thus, the areas with different lattice parameters simultaneously illuminated by the incident beam lead to the peak broadening.

The angle $\gamma$ between the [100]$_{SC}$ and [010]$_{SC}$ axes differs from its average value of $\langle \gamma \rangle = 90 \pm 6°$ in a range of 76° to 105° over the whole supercrystal as shown in **Fig. 4d**. Specifically, we find $\gamma > 90°$ close to the top and bottom corners of the supercrystal and $\gamma < 90°$ close to the left and right corners. Thus, the angle pointing towards the corner of the supercrystal is always obtuse. We further calculate the azimuthal position $\varphi$ of the mean line $M$ between the [100]$_{SC}$ and [010]$_{SC}$ axes. This angle can be interpreted as the azimuthal orientation of the unit cell of the supercrystal. The orientation changes inhomogeneously throughout the superlattice in the range from 72° to 97° as shown in **Fig. 4e**. There is no obvious correlation between the lattice orientation and the spatial position within the sample. Overall, these results suggest that the supercrystal is simple cubic on average, but it exhibits substantial local monoclinic distortions.

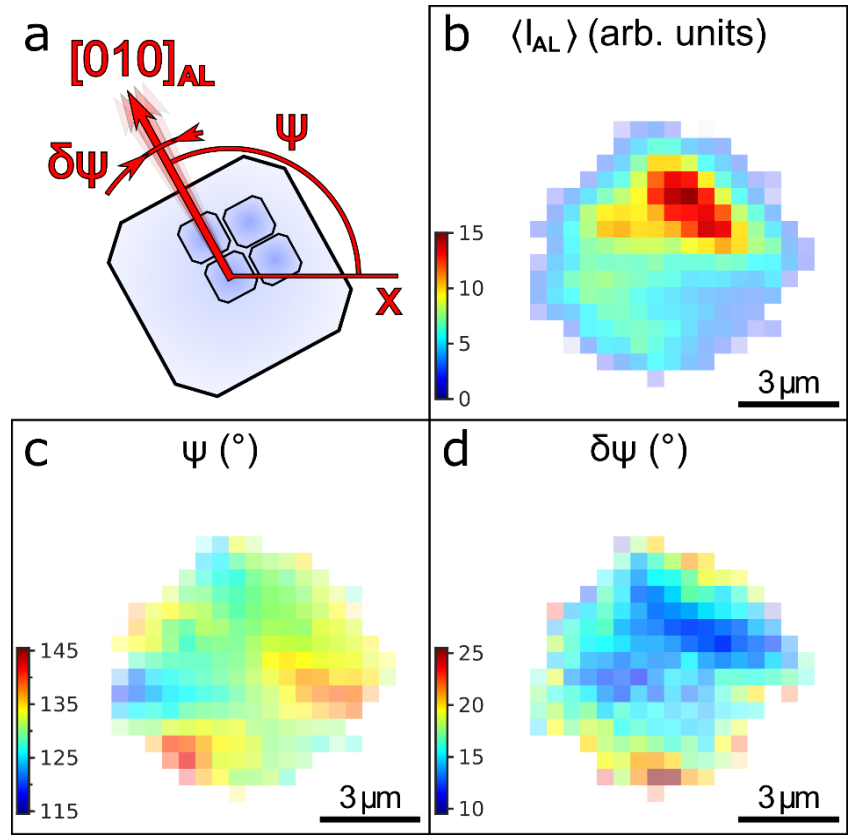

**Figure 5: Spatially resolved WAXS.** (**a**) Definition of the geometrical parameters of the atomic lattice extracted by fitting of the Bragg peaks; (**b**) mean intensity of the WAXS Bragg peaks $\langle I_{AL} \rangle$; (**c**) azimuthal position $\psi$ of the $100_{AL}$ crystallographic axis of the NCs; (**d**) FWHM $\delta\psi$ of the angular disorder of the NCs around the mean azimuthal position $\psi$ extracted from the azimuthal FWHMs of the Bragg peaks by the Williamson-Hall method. The pixel size in (b-d) is 500 nm.

We analyze the Bragg peaks in the WAXS region of individual diffraction patterns at different locations to study the angular orientation of the NCs inside the superlattice. From the WAXS Bragg peak analysis, we extract the average WAXS intensity $\langle I_{AL} \rangle$ and the azimuthal position $\psi$ of the $[010]_{AL}$ axis defined in **Fig. 5a** (see Methods section and Supporting

Information, **Section S6**). In contrast to the intensity of the SAXS Bragg peaks, the WAXS intensity $\langle I_{AL} \rangle$ decreases towards the edges, as shown in **Fig. 5b**, indicating an out-of-plane rotation of the NCs that shifts the Bragg peaks slightly out of the Ewald sphere.[26] We find that $\psi$ changes in a wide range from 120° to 142° as shown in **Fig. 5c**. The map of $\psi$ resembles that of the azimuthal orientation $\varphi$ of the mean line $M$, shown in **Fig. 4e**. The 45° offset between the $[010]_{AL}$ axis and the mean line $M$ indicates the alignment of the $[110]_{AL}$ axis with the mean line $M$ between the $[100]_{SC}$ and $[010]_{SC}$ axes (see Supporting Information, **Figure S21**).

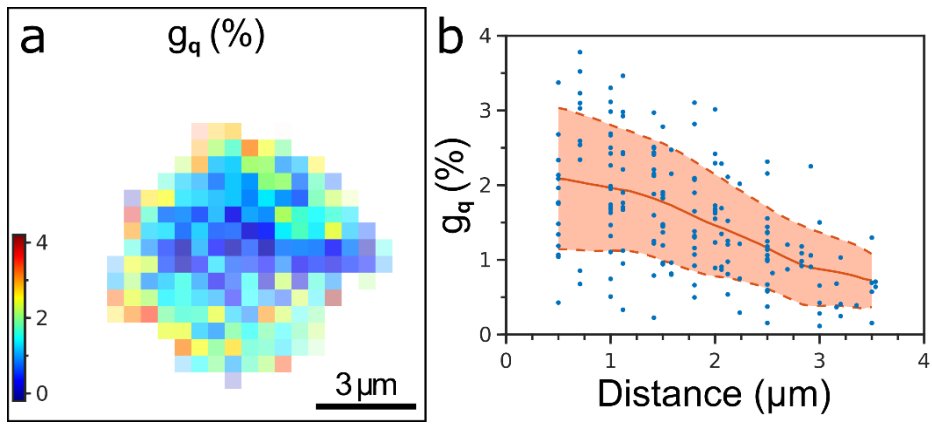

**Figure 6**: **Spatially resolved atomic lattice distortion** (**a**) Atomic lattice distortion $g_q$ extracted from the radial FWHMs of the WAXS Bragg peaks by the Williamson-Hall method. The pixel size is 500 nm. (**b**) The same value $g_q$ for each pixel plotted against the distance from this pixel to the nearest edge of the supercrystal. The red line shows the mean value, the dashed lines indicate the confidence interval of $\pm\sigma$.

From the azimuthal FWHMs of the WAXS Bragg peaks, we extract the angular disorder $\delta\psi$ of the individual nanocrystals at each spatial point by the Williamson-Hall method as shown in **Fig. 5d** (see Methods for details). The disorder is smallest in the center of a supercrystal (9.9°) and increases to a maximum of 24.0° at the edges. The mean value of the angular disorder is $\langle\delta\psi\rangle$ = 16.1±2.8°, which is consistent with previously observed values for similar superstructures.[23,25–27]

Despite the fact that the atomic lattice parameter $a_{AL}$ is constant within the error bars throughout the whole supercrystal (see Supporting Information, **Figure S18**, for the map of $a_{AL}$), we find a difference in the radial width of the Bragg peaks at different locations. By the Williamson-Hall method, we extract the lattice distortion $g_q$ (the ratio $\delta a_{AL}/a_{AL}$, where $\delta a_{AL}$ is the FWHM of the unit cell parameter distribution around the mean value $a_{AL}$) at each spatial point (see Methods section for details). We find a clear trend of increasing atomic lattice distortion towards the edges of the supercrystal with a maximum of 2% at the edge, while it is about 1% at a distance 3 µm into the center, as shown in **Fig. 6**. The trend is even more evident for another supercrystal with particularly good signal-to-noise ratio of the WAXS intensity (see Supporting Information, **Figure S27**).

To rationalize the experimental trend of increased fluorescence energies at the edges of the supercrystal as compared to its center, we carry out density functional modelling of the system. We consider three individual contributions in this regard. First, we recognize that the number of the nearest neighbors at the surface of the supercrystal should be lower than that in the center, leading to stronger exciton confinement and hence increased fluorescence energies at the edges. Indeed, our DFT calculations confirm this trend in **Figure 7a**, which is consistent with the blueshift of the fluorescence spectra observed experimentally for the NCs at the edges. Second, we anticipate that the shorter interparticle spacing (**Fig. 4b**) should facilitate better electronic coupling between the nanocrystals at the edges and, therefore, a decrease in the optical gap at the edges is anticipated. While this expectation is confirmed computationally in **Figure 7b**, we note that it is exactly opposite to what is observed experimentally in **Figures 1b,e** (see Discussion section for details). Third, the supercrystal is compressed at the edges, as evident from **Figure 4b**. While it is reasonable to assume that the compressive strain will mostly manifest in a denser packing of the soft oleylamine/oleic acid ligand sphere of the NCs, we also consider a partial compression of the hard-inorganic lattice-core. In **Figure 7c** we calculate the effect of such compression on the HOMO–LUMO gap ($E_{gap}$) of the NC. While axial stress applied to the CsCl-terminated surface of the CsPbBr$_2$Cl particle results in a steady increase of the optical gap consistent with the experiment, similar stress on the CsBr-terminated surfaces of both particles are found to both increase or decrease $E_{gap}$, depending on the magnitude of the applied stress.

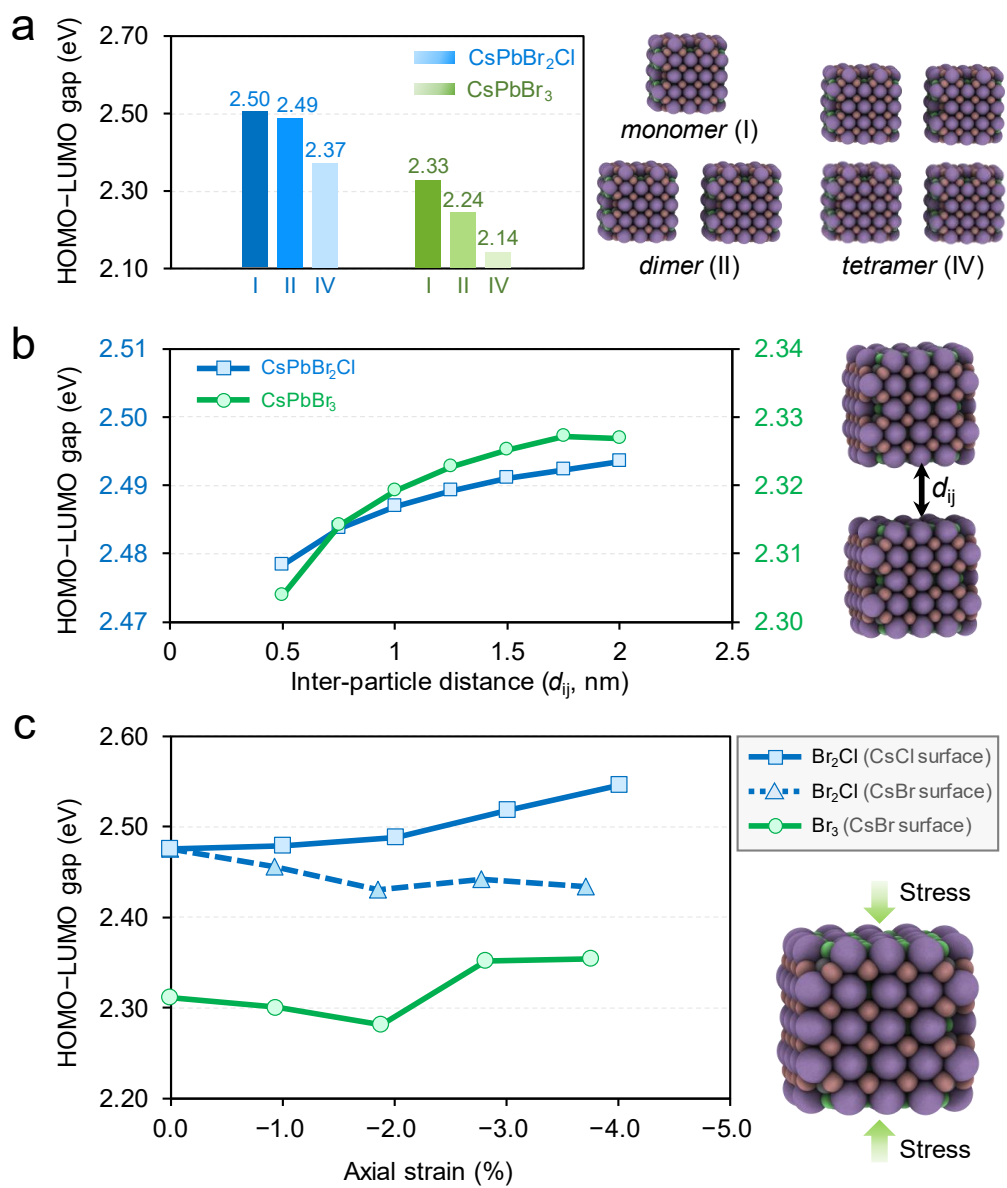

**Figure 7: Density Functional Modelling.** Computed HOMO–LUMO gaps as a function of **(a)** number of neighboring particles considered (dimers and tetramers are 0.5 nm apart), **(b)** distance between two adjacent particles, and **(c)** applied axial strain, for both CsPbBr$_2$Cl and CsPbBr$_3$ particles. All energies are in eV computed at PBE/DZVP level of theory.

Overall, our computational modelling suggests that the spectral blueshift of the fluorescence from the edges of the supercrystal can be rationalized in terms of a dominating effect of a reduced NC coordination number at the edges and – to some extent – compressive atomic lattice strain. Since the experimentally measured spectral shift is seemingly a combination of all three effects discussed above, a fully quantitative prediction would require more detailed knowledge on their relative contributions as well as the relative orientation and positions of individual nanocrystals, which are currently not available.

**Discussion**

When NCs are self-assembled into supercrystals from colloidal solution via slow drying, the increasing curvature and surface tension of the evaporating solvent invokes compressive strain on the supercrystal.[31,32] We hold such a strain responsible for the observed compression of the unit cell parameter by over 20% of the $CsPbBr_2Cl$ NC supercrystals in **Fig. 4b**. This compression is possible due to the softness of the oleylamine/oleic acid ligand shell of the NCs, enabling a large decrease of the interparticle distance by growing interdigitation of adjacent ligand spheres. We note that the compression occurs gradually over a length scale of many lattice planes (>1 µm), meaning that it is not a localized surface reconstruction as commonly observed in atomic crystals.[12] The accompanying loss in angular correlation of the constituting NCs with the superlattice fits to a scenario where strain in the supercrystal is partially relieved by forming local structural defects. The comparison of the average (**Fig. 3c**) *vs.* the local (**Fig. 4**) structure of the supercrystal shows that such distortions are indeed frequently present. We note that recent work on $CsPbBr_3$ NC supercrystals reported perfect structural coherence exclusively in the out-of-plane direction.[22] Since our experiment is only sensitive to in-plane structural features, the findings here are not contradictive to that report.

Our results in **Fig. 4c** support the view of Kapuscinsky et al. that strain during the self-assembly is initially isotropic but later becomes increasingly anisotropic.[31] In a simple cubic supercrystal, the preferred direction for anisotropic structural changes to manifest is the $<111>_{SC}$, which will result in a shear deformation of the ligand spheres.[33] The expected structure of the supercrystal after this shear deformation is reasonably resembled by the local structure depicted in **Fig. 4**.

The compression in the supercrystals is not exclusively limited to the soft ligand sphere. With an interparticle distance of <1 nm close to the edges of a supercrystal, the space for the two ligand spheres of adjacent NCs is so constrained, that the inorganic cores of the NCs become compressed as well (**Fig. 6**). Strain in lead halide perovskite thin films plays an important role for their optoelectronic properties and application in photovoltaic devices.[34] Our fluorescence and fluorescence lifetime data in **Figures 1 and 2** suggests that this is also the case for lead halide NC supercrystals. A comparison of **Figure 1** with **Figure 4b** reveals a strong correlation between the gradual blueshift of the fluorescence peak wavelength and the progressive compression of the supercrystal. We suggest that the shift by up to 20 meV is the result of three, partially competing phenomena: 1) a loss in structural coherence as well as isoorientation of NCs (**Fig. 5c** and **d**) 2) a decrease of the interparticle distance (**Fig. 4b**), and 3) the distortion of the atomic lattices of the NCs (**Fig. 6**). Our DFT calculations in **Figure 7a** suggest that the first effect should be associated with a significant blueshift of the fluorescence due to reduced coupling, consistent with a previous report about the importance of structural coherence for electric transport in supercrystals.[6] While the second effect can only lead to a red shift (**Fig. 7b**), the third effect is also shown to invoke a blueshift for specific facets or magnitudes of strain (**Fig. 7c**).

With reference to several studies on $CsPbBr_3$ NCs which reported a red-shifted fluorescence after assembly into supercrystals, we note that the resultant peak wavelength may further be affected by the concomitant changes in the dielectric environment and photon propagation, aging, miniband formation as well as cooperative emission.[8,9,35,36]. However, most of these observations were made under markedly different conditions, such as low temperature, prolonged exposure to air or self-assembly at the liquid/gas interface, which may be the reason that they are not a dominating factor in our study.

We note a previous report on the spatially resolved fluorescence of $CsPb(I_{0.28}Br_{0.72})_3$ NC supercrystals with a similar fluorescence blueshift between the center and the edge.[24] As a main conclusion, gradual release of $I_2$ gas under intense laser illumination led to the blueshift since lead bromide perovskites exhibit a larger bandgap than the corresponding lead iodide perovskites. The authors argued that the $I_2$ loss commenced from the edges towards the center, which would explain the spatial fluorescence variations. In $CsPbBr_2Cl$ however, this mechanism is not easily applicable since the reduction potential of $Br^-$ is much lower than that of $I^-$. In line with this, $CsPbBr_3$ NC supercrystals without a halide mixture show a similar blueshift, indicating that a change in the mixed halide composition is not required to observe the effects reported here.

The decrease of the fluorescence lifetime in **Figure 2** is also strongly correlated with the gradual compression of the supercrystal towards the edges. Moreover, many supercrystals exhibit particularly decreased lifetime values at the corners, which bears similarities with the anisotropic changes in the lattice spacings in **Figure 4c**, highlighting again the correlation between structural and optical properties. We speculate that the increased atomic lattice distortion and loss of structural coherence near the edges of the supercrystals result in a reduced stability of the excited state of the emitting NCs. This view is supported by the decreased radiative lifetime values from these locations. In view of the currently pursued application of lead halide NC supercrystals as superfluorescent emitters,[8,10] this would imply that bright and coherent emission originates from the center of the supercrystals as long as they are freshly prepared. Conversely, for aged $CsPbBr_2Cl$ NC supercrystals, the lifetimes are longest at the edges, which points to an increased stability of the excited state, potentially due to the formation of a protective oxide shell.[37]

In conclusion, supercrystals of lead halide perovskite NCs self-assembled from solution exhibit a loss in structural coherence, an increasing atomic misalignment between adjacent NCs, and compressive strain near their surfaces. These structural distortions are strongly correlated with a blue-shifted fluorescence and decreased radiative lifetimes. We note that structural distortion and surface defects have been shown to strongly affect the fluorescence properties in *atomic* crystals, such as transition metal dichalcogenides.[13–17] The structure-fluorescence correlations in *supercrystals* revealed here are thus another example for the analogy between atoms and NCs as so-called quasi-atoms.

**Acknowledgement**

We acknowledge DESY (Hamburg, Germany) for the provision of experimental facilities. Parts of this research were carried out at PETRA III synchrotron facility and we would like to thank the beamline staff for assistance in using the Coherence Application P10 beamline. This work was supported by the Helmholtz Associations Initiative Networking Fund (grant No. HRSF-0002), the Russian Science Foundation (grant No. 18-41-06001) and the DFG under grants SCHE1905/8-1, SCHE1905/9-1, AN680/6-1 and SCHR700/38-1. D.L., N.M., D.As., Y.Y.K., M.Sp., and I.A.V. acknowledge support of the project by Edgar Weckert.

**Authors contribution**

D.L., C.K., and J.H. contributed equally to this work. D.L., D.As., J.C., Y.Y.K., N.M., I.Z. and M.Sp. performed the X-ray scattering experiments. C.K., S.W. and J.W. synthesized the NCs, conducted optical absorption and fluorescence measurements in solution and prepared all samples. J.H. carried out the confocal fluorescence (-lifetime) measurements. M.M. performed DFT calculations and A.M. undertook the SEM and AFM measurements. F.S., A.J.M., K.B., D.A., I.A.V. and M.S. conceived and supervised the project. D.L., J.H., M.M., I.A.V. and M.S.

wrote the manuscript with input from all authors. All authors have given approval to the final version of the manuscript.

**Competing interests**

The authors declare no competing interests.

**Data availability**

The datasets generated during and/or analyzed during the current study are available from the corresponding author on reasonable request.

**Methods**

*Chemicals*

1-Octadecene (ODE), technical grade, 90%, Sigma Aldrich; Oleic acid (OA), 97%, Acros Organics; Oleylamine (OAm), 80 - 90%, Acros Organics; Caesium carbonate ($Cs_2CO_3$), 99.99% (trace metal basis), Acros Organics; Lead(II)chloride ($PbCl_2$), 99.999% (trace metal basis), Sigma Aldrich; Lead(II)bromide ($PbBr_2$), ≥98%, Sigma Aldrich; Toluene, 99.8%, extra dry, AcroSeal, Acros; Tetrachloroethylene (TCE), ≥99%, Acros Organics; Kapton® polyimide membranes (125 μm thickness) were purchased from DuPont; Si/SiOx wafers (200 nm SiOx thickness) were purchased from Siegert Wafer GmbH. All chemicals were used as purchased.

*Preparation of Cs-oleate*

203.5 mg $Cs_2CO_3$ (0.625 mmol) was loaded into a 25 mL three-neck flask along with 10 mL 1-octadecene and 0.625 mL oleic acid, dried for 1 h at 120 °C and then heated to 150 °C under nitrogen atmosphere until all $Cs_2CO_3$ reacted with oleic acid. The mixture was kept in a glovebox and heated to 110 °C before injection.

*Synthesis of CsPbX$_3$ nanocrystals*

CsPbX$_3$ NCs were made by a hot-Injection synthesis using a modified literature method.[38] To synthesize 9 nm CsPbBr$_3$ or 7 nm CsPbBr$_2$Cl NCs, 138 mg (0.38 mmol) PbBr$_2$ or 92 mg (0.25 mmol) PbBr$_2$ and 35 mg (0.125 mmol) PbCl$_2$ were degassed in 10 mL ODE in a 25 mL three-neck flask under reduced pressure at 120 °C for 2 h. Then, 1 mL of dried oleylamine (OAm) and 0.5 mL of dried oleic acid (OA) were injected at 120 °C under nitrogen atmosphere with continuous stirring and the reaction mixture was heated to 160 °C. After the solubilization was completed, 0.8 mL of a previously prepared solution of Cs-oleate in ODE (0.125 M) was swiftly injected, and the reaction mixture was cooled to room temperature using an ice-bath.

*Isolation and purification of CsPbX$_3$ nanocrystals*

CsPbX$_3$ NCs were collected by centrifuging the suspension (7000 rpm, 10 min), decanting the supernatant, and collecting the precipitate. The precipitate was centrifuged again without addition of a solvent (7000 rpm, 5 min), and the resulting supernatant was removed with a syringe, to separate the traces of residual supernatant. The precipitate was dissolved in 2 mL hexane and centrifuged again (2500 rpm, 5 min) to remove aggregates and larger particles. The resulting supernatant was filtered through a 0.2 µm PTFE syringe filter and stored as stock solution inside of a glovebox with a typically concentration of 16 mM following Maes *et al.*[39]

*Self-assembly of NC superlattices*

For the growth of supercrystals, different substrates (Si wafer, Kapton, glass) were used, depending on the desired experiment. The self-assembly experiment was set up in a glass Petri dish (with a 60 mm diameter), for this purpose three substrates each were placed in such a Petri dish together with a PTFE-lid filled with 1 mL tetrachloroethylene. To each of these substrates, 40 µL of a 1 - 3 mM solution of the perovskites in TCE was added. The lid of the Petri dish was closed, covered with aluminum foil, and allowed to stand for 24 h. After that, the lid was opened and left for another 5 h to dry completely. All self-assembly preparations were performed under inert atmosphere. The more monodisperse the size distribution of the perovskites, the better the resulting superlattices

*Spatially resolved optical measurements*

All spatially resolved optical measurements were performed using a home-built inverted confocal laser scanning microscope. The measurements were performed on glass substrates utilizing a high numerical aperture oil immersion objective (NA = 1.4) and a 405 nm pulsed diode laser (Picoquant LDH P-C-405) with variable repetition rates (Picoquant PDL 800-D laser driver) as the excitation source. Under these conditions the lateral resolution of the instrument is approximately 200 nm. A single photon avalanche diode (MPD PDM Series) was used in conjunction with the Picoquant HydraHarp 400 as a time-correlated single photon counting system to detect time-resolved fluorescence. Time-resolved data acquisition and analysis was performed using Picoquants SymPhoTime 64 software package. The spectral data was recorded using an Acton Spectra Pro 2300i spectrometer with a 300 grooves/mm grating. The detector temperature (Princeton PIXIS CCD) was kept steady at -45 °C.

*X-ray diffraction experiment*

The nano-diffraction experiment was performed at the Coherence Applications beamline P10 of the PETRA III synchrotron source at DESY. An X-ray beam with the wavelength $\lambda = 0.0898$ nm ($E = 13.8$ keV) was focused down to a spot size of approximately $400 \times 400$ nm$^2$ (FWHM) with a focal depth of about 0.5 mm at the GINIX nano-diffraction endstation.[40] The two-dimensional detector EIGER X 4M (Dectris) with $2070 \times 2167$ pixels and a pixel size of $75 \times 75$ μm$^2$ was positioned 412 mm downstream from the sample. The detector was aligned ~6 cm off-centre in both directions normal to the incident beam to allow simultaneous detection of SAXS and WAXS. We performed a spatially resolved scan of the sample on a Kapton substrate by $25 \times 25$ spatial points with 500 nm step size and collected 625 diffraction patterns in transmission geometry. The exposure time was set to 0.5 s to prevent radiation damage of the sample. The background scattering pattern from a pure Kapton film was subtracted from every collected pattern.

*Bragg peak analysis*

Each diffraction pattern was interpolated onto a polar coordinate grid with the origin at the direct beam position. The radial profiles were obtained by averaging along the azimuthal coordinate. To extract parameters of the WAXS and SAXS Bragg peaks separately, we fitted each of them by the 2D Gaussian function

$$I(q, \varphi) = \frac{I_0}{2\pi\sigma_q\sigma_\varphi} \exp\left[-\frac{(q-q_0)^2}{2\sigma_q^2} - \frac{(\varphi-\varphi_0)^2}{2\sigma_\varphi^2}\right],$$

where $I_0$ is the integrated intensity, $q_0$ and $\varphi_0$ are the radial and azimuthal central positions, and $\sigma_q$ and $\sigma_\varphi$ are the corresponding root mean square (rms) values. The FWHMs of the Bragg peaks were evaluated according to relations: $w_q = 2\sqrt{2ln2}\sigma_q$ and $w_\varphi = 2\sqrt{2ln2}\sigma_\varphi$. The fitting was done in the appropriate region of the polar coordinates with a single isolated Bragg peak.

For the SAXS peaks, the parameters were pairwise averaged for the corresponding Friedel pairs of the Bragg peaks to improve statistics. The resulting momentum transfer values and angles were used to calculate the real space parameters of the unit cell: the length of the basis vectors $a_1$ and $a_2$, the angle $\gamma$ between them and the average azimuthal position $\varphi$ counted counterclockwise from an arbitrary horizontal axis (see Supplementary Materials for details).

For the WAXS peaks, we calculated an average Bragg peak intensity $I_{AL}$ and the azimuthal position $\psi$ of the $[010]_{AL}$ axis. To obtain the average azimuthal position $\psi$, we averaged all four azimuthal positions for $010_{AL}$, $020_{AL}$, $110_{AL}$, and $\bar{1}00_{AL}$ Bragg peaks, but corrected the last two values by +45° and -45°, respectively. We used the Williamson-Hall method[41] to analyze the size of the WAXS Bragg peaks at each spatial point of the supercrystal. The FWHM of the Bragg peak is determined by the NC size and the lattice distortion as follows:

$$w_{q,\varphi}^2(q) = \left(\frac{2\pi K}{L}\right)^2 + \left(g_{q,\varphi} q\right)^2, \qquad (1)$$

where $w_{q,\varphi}$ is the FWHM of the Bragg peak at $q$ in radial or azimuthal direction, respectively, $L$ is the NC size, $g_{q,\varphi}$ is the radial or angular lattice distortion of the atomic lattice, respectively, $K$ is a dimensionless shape parameter that is about 0.86 for cubic NCs.[42] The radial lattice distortion $g_q$ calculated from the radial FWHM $w_q$ is equal to the ratio $\delta a_{AL}/a_{AL}$, where $\delta a_{AL}$ is the FWHM of the unit cell parameter distribution around the mean value $a_{AL}$. The angular lattice distortion $g_\varphi$ calculated from the azimuthal FWHM $w_\varphi$ is equal to the FWHM $\delta\psi$ of the angular distribution of the NCs around their average azimuthal position $\psi$. For the spatially resolved analysis of the FWHMs, the NC size $L$ was fixed at the value, obtained from the average radial profiles. For details of the analysis, see Supplementary materials.

*Scanning electron and atomic force microscopy*

SEM imaging of supercrystals on Si/SiO$_x$ devices was performed with a HITACHI model SU8030 at 30 kV. To estimate the thickness of micro-crystals, samples were titled by 45° with respect to the incoming electron beam. AFM investigations were conducted with a Bruker MultiMode 8 HR in contact mode.

*Density functional theory calculations*

All computations are performed using the CP2K 5.1 program suite using the Quickstep module.[43] The PBE exchange correlation functional,[44] a dual basis of localized Gaussians and plane waves (GPW)[45] with a 350 Ry plane-wave cutoff, double-ζ basis-set augmented with polarization functions (MOLOPT variant),[46] and GTH pseudopotentials[47] for core electrons are used for all calculations. The van der Waals (VDW) interaction was accounted for by employing Grimme's DFT-D3 method.[48] SCF convergence criterion was set at $10^{-6}$ for all calculations.

Initial geometries of CsPbX$_3$ (X = Cl, Br) nanocrystals were obtained by cutting small cubes (~2.4 nm) from the bulk, exposing the CsX layer at the surface and maintaining overall charge neutrality of the particle.[49] All structures were then optimized in vacuum using the BFGS optimizer imposing non-periodic boundary conditions with a wavelet Poisson solver,[50] setting a maximum force of 5 meV·Å$^{-1}$ ($10^{-4}$ hartree/bohr) as convergence criteria. For these non-periodic systems, axial strain was simulated by fixing the length of one side of the cube. If the relaxed cubic nanocrystal has side length $a \times b \times c$, and stress is to be applied along the Z-direction, "c" is fixed at some c′ by constraining the z coordinates of both the top and bottom surface-atoms along the z-direction, with all other coordinates of all atoms relaxed. % Strain is reported as (c′ − c)/c × 100%. For calculations involving dimers and tetramers, 2/4 monomers were explicitly considered, but *periodic* boundary condition was imposed with at least 10 Å vacuum above the surface of the nanocluster to avoid spurious interaction with its periodic image.

# Spatially resolved fluorescence of caesium lead halide perovskite supercrystals reveals quasi-atomic behavior of nanocrystals

Dmitry Lapkin[1,*], Christopher Kirsch[2,*], Jonas Hiller[2,*], Denis Andrienko[3], Dameli Assalauova[1], Kai Braun[2], Jerome Carnis[1], Young Yong Kim[1], Mukunda Mandal[3], Andre Maier[2,4], Alfred J. Meixner[2,4], Nastasia Mukharamova[1], Marcus Scheele[2,4,+], Frank Schreiber[4,5], Michael Sprung[1], Jan Wahl[2], Sophia Westendorf[2], Ivan A. Zaluzhnyy[5], Ivan A. Vartanyants[1,6,+]

1. Deutsches Elektronen-Synchrotron DESY, Notkestraße 85, D-22607 Hamburg, Germany

2. Institut für Physikalische und Theoretische Chemie, Universität Tübingen, Auf der Morgenstelle 18, D-72076 Tübingen, Germany

3. Max Planck Institute for Polymer Research, Ackermannweg 10, D-55128 Mainz, Germany

4. Center for Light-Matter Interaction, Sensors & Analytics LISA[+], Universität Tübingen, Auf der Morgenstelle 15, D-72076 Tübingen, Germany

5. Institut für Angewandte Physik, Universität Tübingen, Auf der Morgenstelle 10, D-72076 Tübingen, Germany

6. National Research Nuclear University MEPhI (Moscow Engineering Physics Institute), Kashirskoe shosse 31, 115409 Moscow, Russia

* These authors contributed equally

+To whom correspondence should be addressed

## Section S1. Optical absorption and fluorescence of CsPbBr$_2$Cl and CsPbBr$_3$ nanocrystals in solution

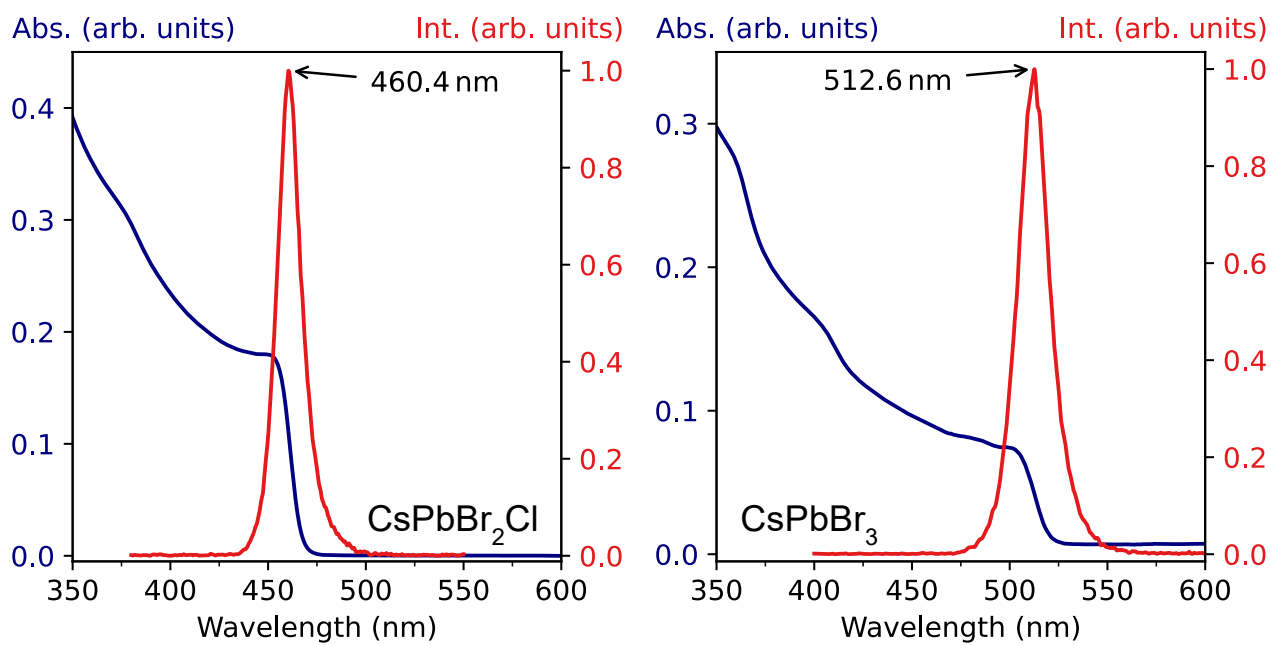

**Figure S1.** (a) Absorption and emission spectrum of CsPbBr$_2$Cl NPs dispersed in Toluene. (b) Absorption and emission spectrum of CsPbBr$_3$ NPs dispersed in Hexane.

Optical measurements were performed on a UV-vis-NIR spectrometer (Agilent Technologies, Cary 5000) and a fluorescence spectrometer (PerkinElmer FL8500). All spectra were acquired under ambient conditions in Toluol at room temperature (25 °C) in a cuvette of 1 cm pathlength.

### Section S2. Fluorescence lifetime imaging

A fast fluorescence lifetime imaging microscopy (fast FLIM) image is recorded by scanning the excitation laser over an area of interest while recording the time-resolved fluorescence in the form of a time-correlated single photon counting (TCSPC) histogram at each pixel. The fast lifetime hereby obtained for each pixel is to be understood as the mean photon arrival time after the excitation laser pulse and therefore includes the time the light takes to travel through the instrument. The histogram of the excitation pulse is called the instrument response function (IRF) and was directly measured by the means of scattered light from a clean glass substrate. The physical decay of the sample is obtained by fitting the experimental decay curves recorded at each pixel by employing the IRF in an n-exponential reconvolution with a maximum likelihood estimation method employed for fit optimization.

For both compositions of the investigated self-assembled supercrystals, good results were obtained by fitting a monoexponential reconvolution to the experimental decay curves. Exemplary fits for the center pixels in the FLIM images of the self-assembled supercrystals composed of $CsPbBr_2Cl$ NCs and $CsPbBr_3$ NCs are depicted in **Fig. S2** and **Fig. S3** respectively.

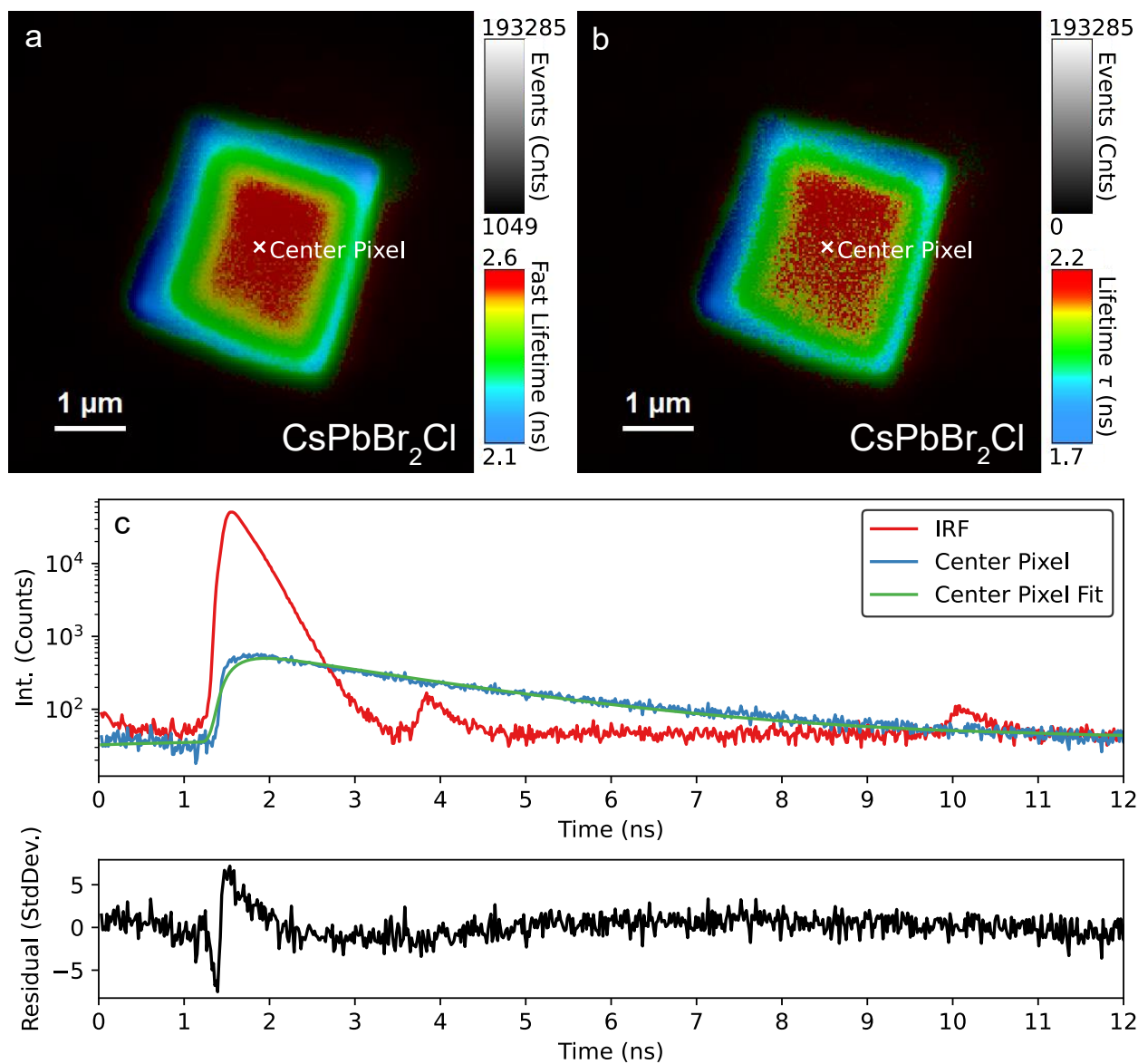

**Figure S2.** For each pixel of the FLIM images the fluorescence intensity is encoded on a brightness scale while fast lifetimes and fluorescence lifetimes are displayed in RGB false color. (a) Fast FLIM image of a self-assembled CsPbBr$_2$Cl supercrystal and (b) the corresponding fitted FLIM image obtained through a pixel-by-pixel monoexponential reconvolution. (c) The experimentally acquired IRF as well as an exemplary decay curve of a pixel in the center of the supercrystal and its corresponding monoexponential fit.

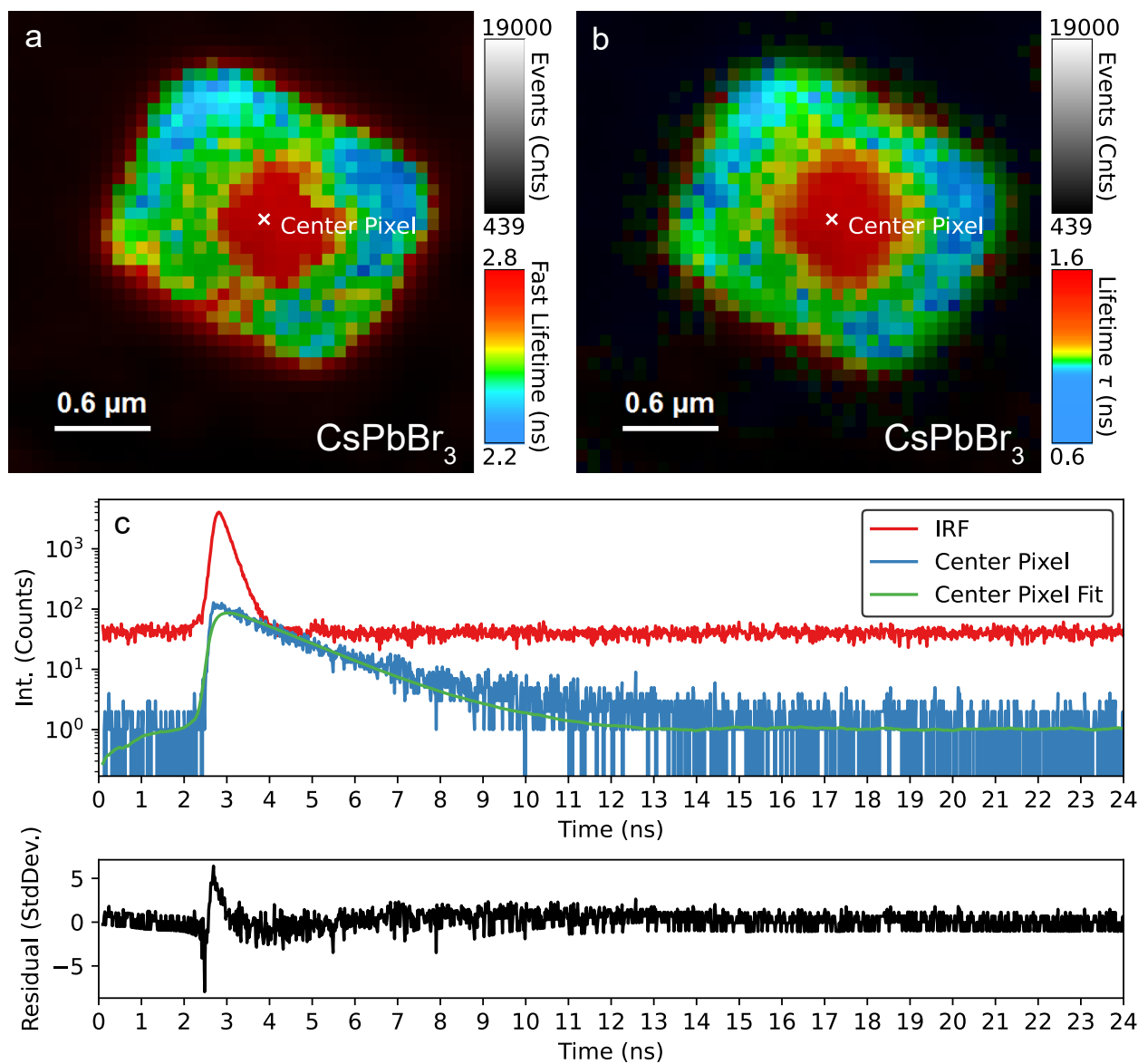

**Figure S3.** For each pixel of the FLIM images the fluorescence intensity is encoded on a brightness scale while fast lifetimes and fluorescence lifetimes are displayed in RGB false color. (a) Fast FLIM image of a self-assembled $CsPbBr_3$ supercrystal and (b) the corresponding fitted FLIM image obtained through a pixel by pixel monoexponential reconvolution. (c) The experimentally acquired IRF as well as an exemplary decay curve of a pixel in the center of the supercrystal and its corresponding monoexponential fit.

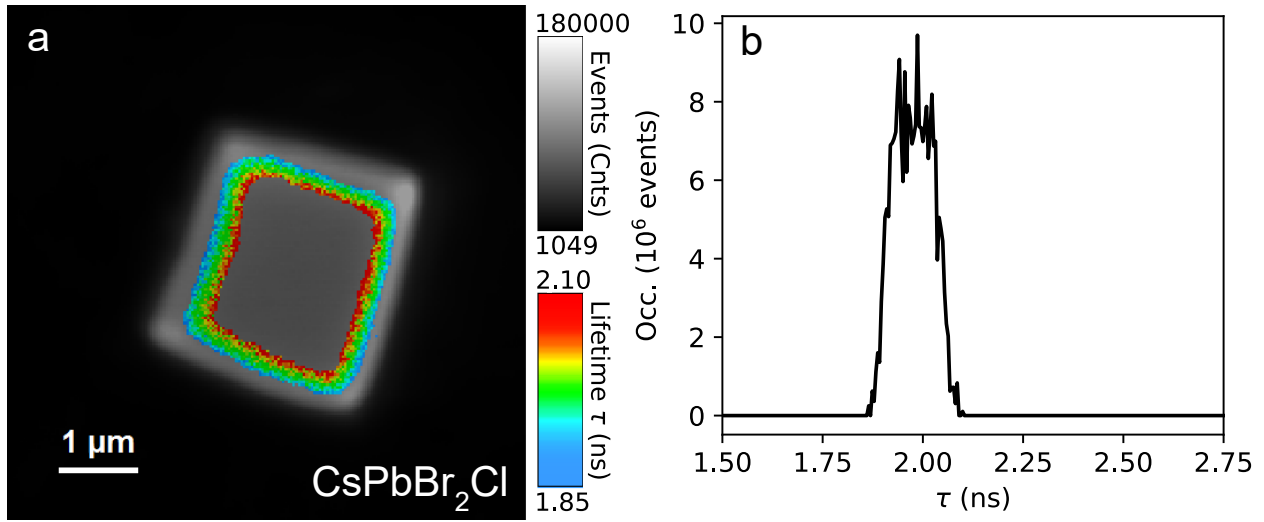

**Figure S4.** (a) Region of interest (ROI) FLIM image of the intermediate area of the CsPbBr$_2$Cl supercrystal depicted in Fig. S2. (b) The occurrence of the $\tau$-values associated with the monoexponential decay throughout the ROI FLIM image. This ROI FLIM image demonstrates that the decrease of the $\tau$-values when scanning from the center of the supercrystal towards its edges is rather gradual. The "step-like" decrease that **Fig. S2** seems to imply is due to the chosen RGB color-scaling.

### Section S2.1. Determining the shortest euclidean distance to the supercrystal edge for each pixel

The first step in the analysis is differentiating between the pixels located inside and those located outside of the supercrystal. This was achieved by assigning the lifetime value 0 to each pixel for which the measured fluorescence intensity is less than half the maximum recorded fluorescence intensity. Because the transition between the inside and the outside of the supercrystal is characterized by a steep drop in the recorded fluorescence intensity, all pixels inside the supercrystal are unaffected and keep their non-zero lifetime value, while all pixels outside of the supercrystal are assigned the lifetime value 0. The masking process for the CsPbBr$_2$Cl supercrystal is depicted in **Fig. S5**. The pixel-to-edge distance for all pixels inside the supercrystal is then the shortest distance between a pixel with a non-zero lifetime value and a pixel with the associated lifetime value 0. The free Python machine learning library Scikit-learn was employed for the actual calculations.

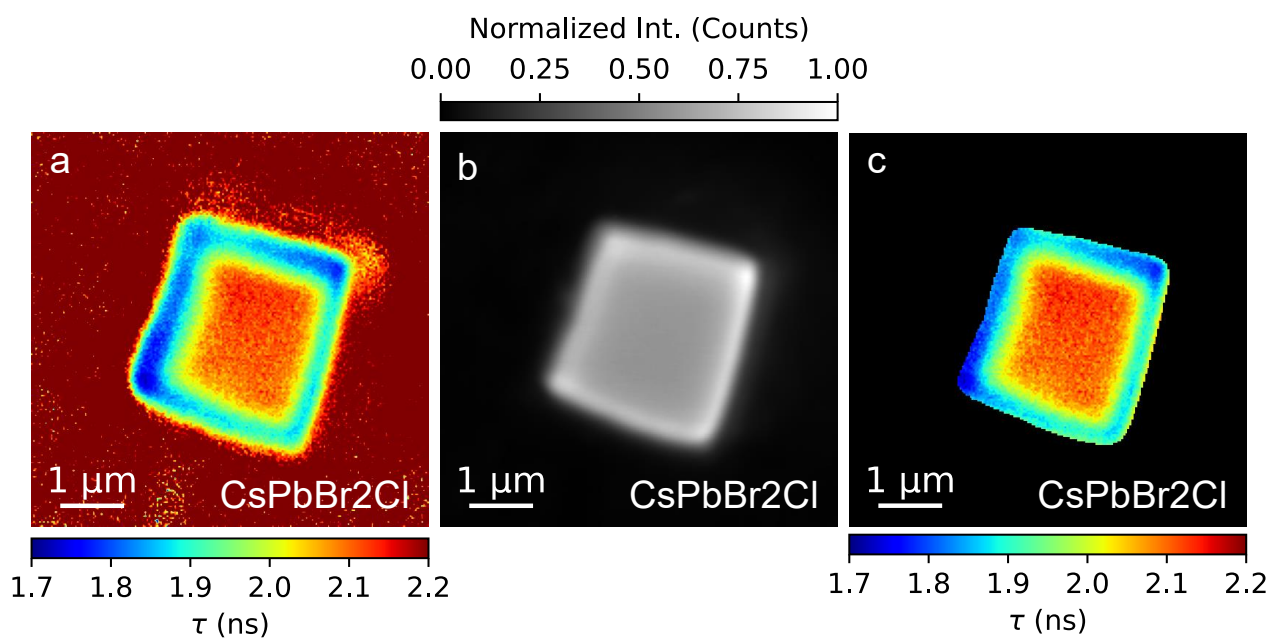

**Figure S5.** For each of the 200x200 pixels the fluorescence lifetime (**a**) as well as the fluorescence intensity (**b**) are recorded. (**c**) Result of assigning the lifetime value 0 to all pixels for which the recorded fluorescence intensity is less than half the maximum recorded fluorescence intensity. All pixels inside the supercrystal remain unaffected and keep their non-zero lifetime value as determined by the measurement.

## Section S3. Electron microscopy of CsPbBr₂Cl supercrystals

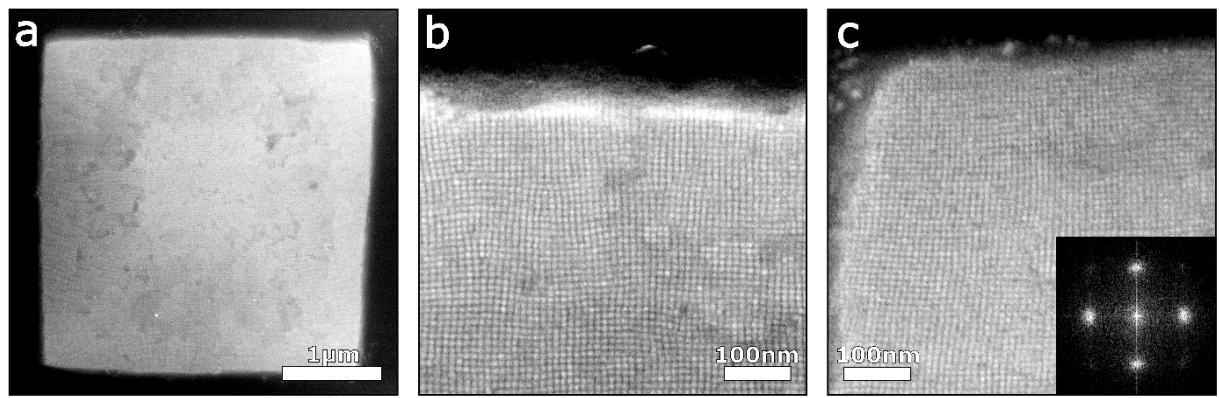

**Figure S6: a**) SEM micrograph of self-assembled supercrystals of CsPbBr$_2$Cl NCs. The NC diameter is rather uniform over the whole crystal (7.3±0.4 nm), as indicated by the high-resolution micrograph of an edge **b**) and a corner **c**). The inset of (c) corresponds to the FFT of the corresponding micrograph, indicating a homogeneous four-fold symmetry of the NC arrangement.

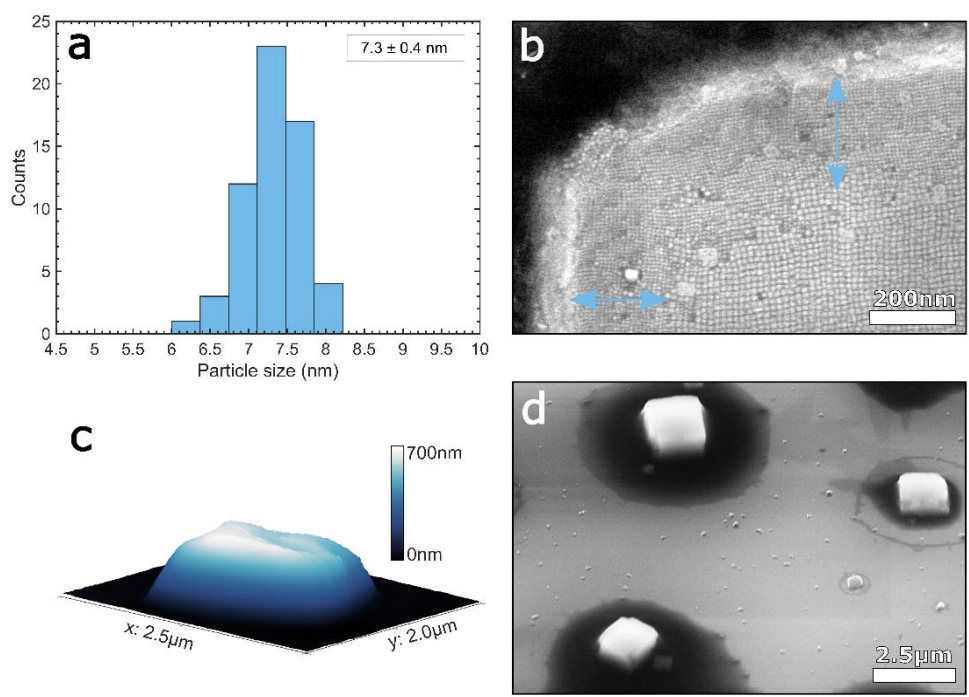

**Figure S7: a**) Distribution of the diameter of the CsPbBr$_2$Cl NCs, measured by SEM. The mean value is 7.3±0.4 nm (size distribution of ~5%). **b**) SEM micrograph of the corner of a less faceted supercrystals occasionally featuring NCs of smaller size. The spatial extent of this subpopulation is limited to ~200 nm from the edges, indicated by the blue arrows. **c**) 3D AFM map of a supercrystal on a Kapton membrane. **d**) SEM micrograph of three supercrystals on a Si/SiOx wafer under a view angle of 45°. Typical supercrystal thicknesses of 580±120 nm can be observed.

**Section S4. Atomic structure of the constituting CsPbBr$_2$Cl NCs**

The azimuthally averaged WAXS intensity calculated for the mean pattern for a supercrystal is shown in **Fig. S8**. There are three prominent peaks originating from the atomic lattice of the CsPbBr$_2$Cl NCs. The peaks can be attributed to 100$_{AL}$, 110$_{AL}$ and 200$_{AL}$ reflections of a cubic superlattice. The peaks at $q \sim 24.5$ nm$^{-1}$ can be attributed to 210$_{AL}$ reflection which is only partially covered by the detector. The structured background in the range of $q = 17$-20 nm$^{-1}$ is from the Kapton film. It cannot be fully subtracted due to the anisotropy of scattering from the Kapton film in different spatial point of the scanned area. The peaks were simultaneously fitted by three Gaussian functions:

$$I(q) = \sum_{i=1}^{3} \frac{I_i}{\sqrt{2\pi\sigma_i^2}} \exp\left[-\frac{(q-q_i)^2}{2\sigma_i^2}\right], \quad \text{(Eq. S1)}$$

where $I_i$ is the integrated intensity, $q_i$ is the momentum transfer and $w_i = 2\sqrt{2ln2}\sigma_i$ is the FWHM of the $i$-th Bragg peak. The peaks are at $q_{100} = 10.930\pm0.005$ nm$^{-1}$, $q_{110} = 15.442\pm0.002$ nm$^{-1}$ and $q_{200} = 21.900\pm0.003$ nm$^{-1}$ giving the unit cell parameter $a_{AL} = 0.574\pm0.001$ nm. The errorbars are the fitting errors. One should note that the expected atomic lattice structure for cesium lead halide perovskites at room temperature is orthorhombic.[1] However, the deviation of the unit cell parameters of such orthorhombic lattice from a cubic lattice is < 2%. We are not able to resolve the peaks of the same order with so small separation due to the size-dependent Scherrer broadening of the Bragg peaks. Thus, we used the pseudocubic indexing of the Bragg peaks, where the 100$_{AL}$ index corresponds to 110$_{AL}$ and 002$_{AL}$ reflections of the orthorhombic structure, the 110$_{AL}$ – to 200$_{AL}$ and 112$_{AL}$, the 200$_{AL}$ – to 220$_{AL}$ and 004$_{AL}$.

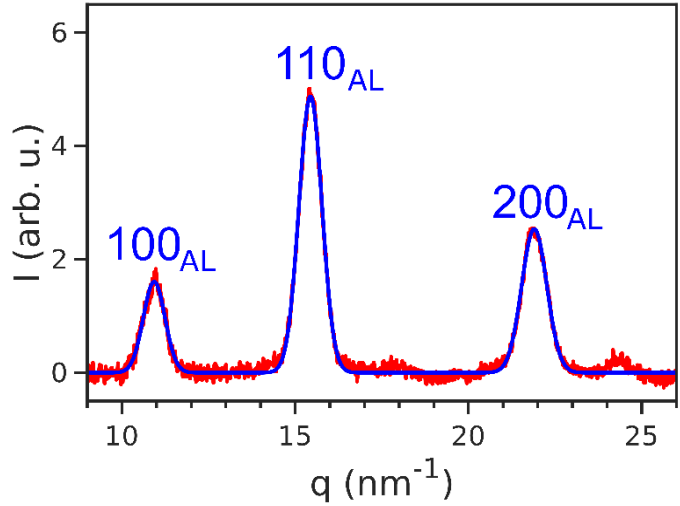

**Figure S8.** Azimuthally averaged intensity profile in WAXS region (red line) and Gaussian fitting including three peaks (blue line). The peaks are indexed according to a pseudocubic structure.

The FWHMs of the peaks extracted by the fitting are $w_{100} = 0.736 \pm 0.011$ nm$^{-1}$, $w_{110} = 0.784 \pm 0.004$ nm$^{-1}$, $w_{200} = 0.856 \pm 0.08$ nm$^{-1}$. We analyzed the FWHMs of the peaks by the Williamson-Hall method.[2] According to the method, the FWHMs of the Bragg peaks are defined by two factors: the size of coherently scattering domain $L$ and the lattice distortion $g$ (the ratio $\delta a_{AL}/a_{AL}$ of the FWHM $\delta a_{AL}$ of the unit cell parameter distribution around the mean value $a_{AL}$). If we assume that the coherently scattering domain is a NC, the Williamson-Hall equation can be written as follows:

$$w^2(q) = \left(\frac{2\pi K}{L}\right)^2 + (gq)^2, \qquad \text{(Eq. S2)}$$

where $w(q)$ is the FWHM of the Bragg peak at momentum transfer $q$, $K$ – a dimensionless shape factor, $L$ – the NC size and $g$ – the lattice distortion.

The first term is the pure Scherrer broadening, where the shape factor K is about 0.85 for the reflections of low orders for a cubic crystallite.[3] One should note, the size of coherently scattering domain can be bigger than a single NC if there are perfectly aligned NCs scattering coherently to the same direction. However, the fact that the WAXS Bragg peaks are much broader in the azimuthal direction than in the radial (see the main text for the WAXS pattern), indicate a high degree of angular disorder of the NCs (will be discussed below) leading to low probability of such a scenario.

To extract the NC size $L$ and the lattice distortion $g$, we fitted the experimentally obtained FWHMs and q-values for the present $100_{AL}$, $110_{AL}$ and $200_{AL}$ peaks with **Eq. S2** as shown in **Fig. S9**. The resulting parameters are $L = 6.8\pm0.1$ nm and $g = 2.3\pm0.1\%$. The resulting NC size $L$ is even smaller than the size $L_{SEM} = 7.3\pm0.4$ nm obtained from SEM measurement that indicates the correctness of the assumption that the coherently scattering domain consist of a single NC. The smaller size can be explained by the lattice twinning inside the NC leading to smaller domains and by the limits of the method.

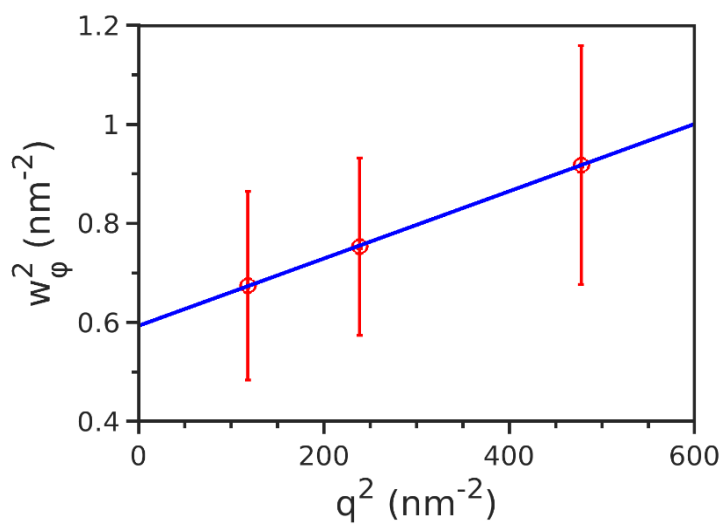

**Figure S9.** Williamson-Hall plot for the radial FWHM values of the WAXS Bragg peaks. The red points are experimental values, the blue straight line is the best fit.

**Section S5. Local superlattice structure**

The single SAXS diffraction patterns at different spatial points of the supercrystal are quite different from the average diffraction pattern shown in **Fig. 3c** in the main text. Examples of the single patterns are shown in **Fig. S10**. Clearly, the Bragg peaks do not maintain their positions in both radial and azimuthal directions and change their shape from point to point.

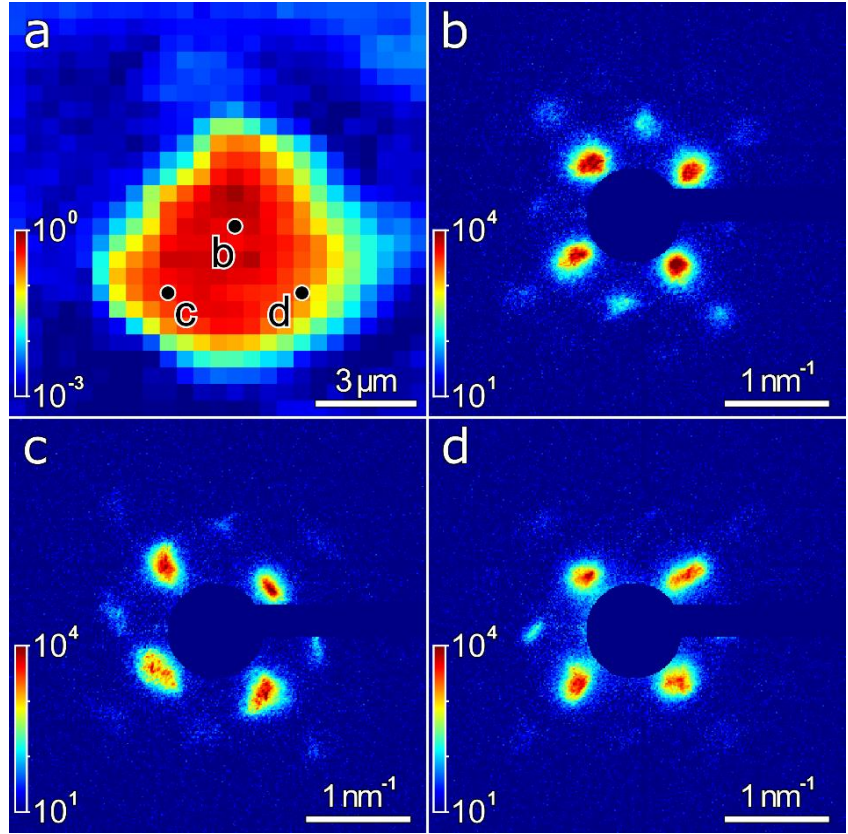

**Figure S10. a)** SAXS intensity-based map of the sample. The pixel size is 500 nm. **b-d)** Examples of single SAXS diffractions patterns collected at the points indicated in panel a).

To study dependence of their parameters on the spatial position within the sample, we evaluated each single diffraction pattern separately. We fitted each of the first order Bragg peaks in the SAXS region by the Gaussian function:

$$I(q, \varphi) = \frac{I_0}{2\pi\sigma_q\sigma_\omega} \exp\left[-\frac{(q-q_0)^2}{2\sigma_q^2} - \frac{(\omega-\omega_0)^2}{2\sigma_\omega^2}\right], \quad \text{(Eq. S3)}$$

where $I_0$ is the integrated intensity, $q_0$ and $\omega_0$ are the radial and azimuthal positions, $w_q = 2\sqrt{2ln2}\sigma_q$ and $w_\omega = 2\sqrt{2ln2}\sigma_\omega$ are the corresponding FWHMs of the Bragg peak. The fitting was done in the appropriate region of the polar coordinates with single isolated Bragg peak.

There are two Friedel pairs of the Bragg peaks corresponding to reflections from the $(100)_{SL}$ and $(010)_{SL}$ superlattice planes. Counting from the right-pointing horizontal axis, the 1st and 3rd peaks correspond to the $(100)_{SL}$ plane and the 2nd and 4th – to the $(010)_{SL}$ plane. We averaged the intensities, the radial and azimuthal positions and the FWHMs within each pair to get more reliable characteristics of the superlattice planes. The azimuthal coordinates of the 3rd and 4th peaks were corrected by -180° prior to the averaging. Finally, we have two sets of the characteristics defined in **Fig. S11a**. The reciprocal space coordinated were converted into the coordinates of the real-space basis vectors $a_1$ and $a_2$.

The azimuthal positions are converted into real space as follows:

$$\varphi_1 = \omega_2 - 90° \quad \text{(Eq. S4)}$$

$$\varphi_2 = \omega_1 + 90° \quad \text{(Eq. S5)}$$

The nearest-neighbor distances $a_1$ and $a_2$ are calculated as:

$$a_1 = \frac{2\pi}{q_1 \cdot \sin\gamma} \quad \text{(Eq. S6)}$$

$$a_2 = \frac{2\pi}{q_2 \cdot \sin\gamma} \quad \text{(Eq. S7)}$$

We also used additional azimuthal coordinates that are the azimuthal position of the mean line $M$ between the $a_1$ and $a_2$:

$$\varphi = \frac{\varphi_1 + \varphi_2}{2} \quad \text{(Eq. S8)}$$

and the angle $\gamma$ between the real space basis vectors $a_1$ and $a_2$, calculated as:

$$\gamma = \varphi_2 - \varphi_1 \quad \text{(Eq. S9)}$$

The definition of the real space coordinates is shown in **Fig. S11b**.

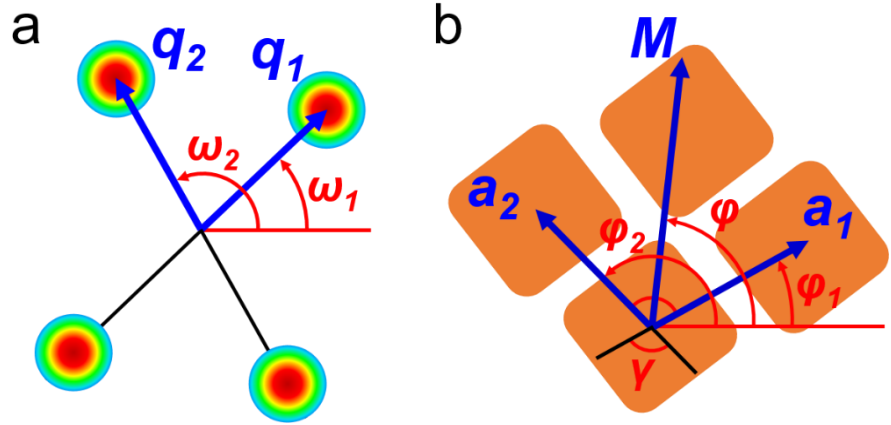

**Figure S11.** a) Scheme of the SAXS diffraction pattern from the superlattice. Only the first order Bragg peaks are shown. Two pairs of Bragg peaks are at momentum transfer values $q_1$ and $q_2$ and azimuthal positions $\omega_1$ and $\omega_2$, respectively. b) Scheme of the real space unit cell of the superlattice. The nearest neighbors are at distances $a_1$ and $a_2$ with azimuthal positions $\varphi_1$ and $\varphi_2$, respectively. The mean line $M$ between $a_1$ and $a_2$ is at azimuthal position $\varphi$. The angle between $a_1$ and $a_2$ is equal to $\gamma$.

The extracted intensities $I_1$ and $I_2$ of the Bragg peaks are shown in **Fig. S12a,d**. The scattering areas basically coincide except the upper and right corners. Vanishing of the 1$^{st}$ Bragg peak possibly indicates out-of-plane rotation of the SL on these supercrystal edges as soon as such rotation bring the peak out of the Ewald sphere.

The extracted momentum transfers $q_1$ and $q_2$ associated with the superlattice plane spacings are shown in **Fig. S12b,e**. As it is clear from the figure, the momentum transfers grow on the edges of the supercrystal. It indicates contraction of the superlattice that is thoroughly discussed in the main text as well as its anisotropy.

The extracted azimuthal positions $\omega_1$ and $\omega_2$ are shown in **Fig. S12c,f**. They clearly indicate rotation of the superlattice around the incident beam (normal to the substrate). The rotation is also discussed in the main text.

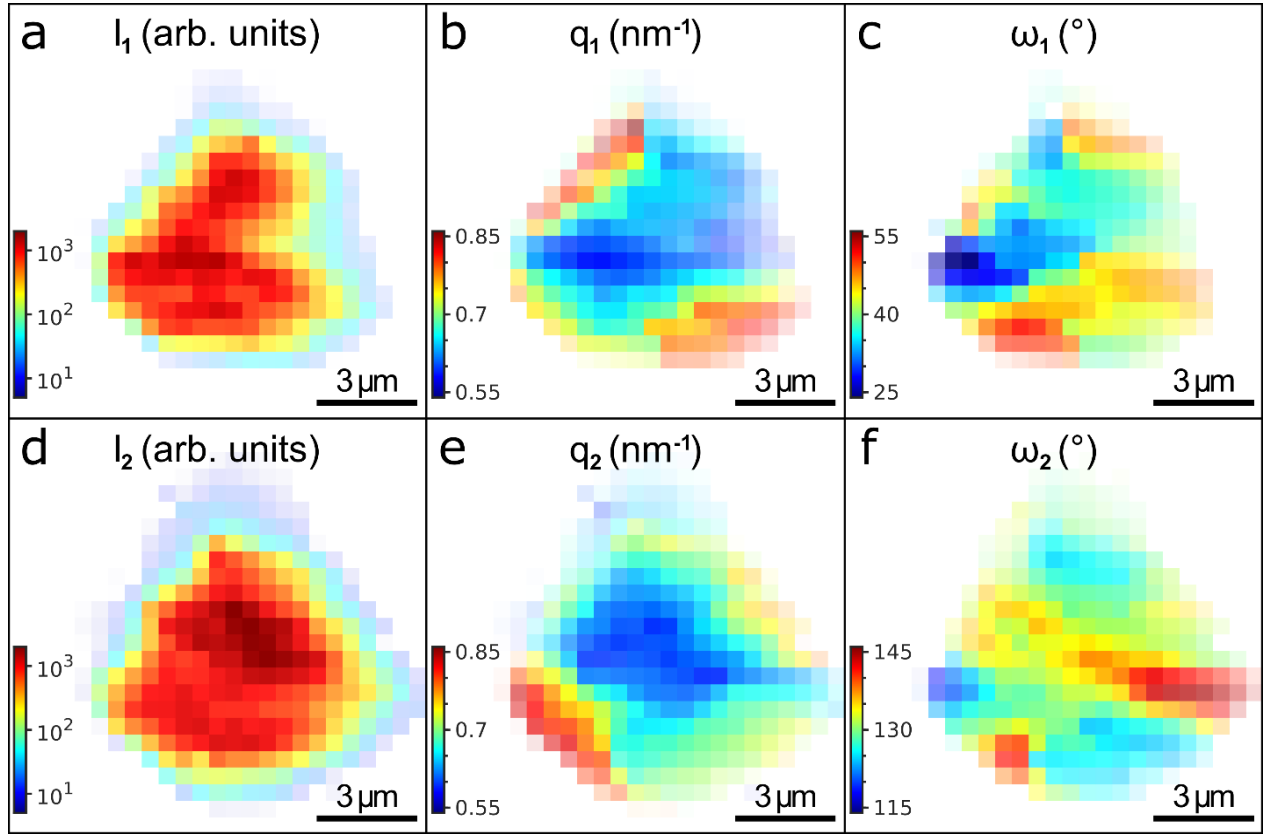

**Figure S12.** Mean extracted parameters: (a, d) intensities, (b, e) momentum transfer values and (c, f) azimuthal positions of (a-c) the 1st and 3rd and (d-f) the 2nd and 4th Bragg peaks. The azimuthal positions are counted counterclockwise from a horizontal axis pointing to the right. The azimuthal position of the 3rd and 4th were corrected by -180° before averaging with their counterparts. The pixel size is 500 nm.

The extracted FWHMs in the radial direction $w_{q1}$ and $w_{q2}$ are shown in **Fig. S13a,c**. Interestingly, in contrast to the momentum transfer values $q_1$ and $q_2$, the main deviations in FWHMs happen on the edges to which the crystallographic axes are perpendicular. This means lower dispersion in the superlattice plane separation normal to the edge of the supercrystal.

The extracted FWHMs in the azimuthal direction $w_{\omega 1}$ and $w_{\omega 2}$ are shown in **Fig. S13b,d**. There is no clear trend in the behavior of these parameters in respect to the spatial position inside the supercrystal. But most of the point having high FWHM values are located in the middle of the supercrystal, indicating higher dispersion of the angle $\gamma$ between the lattice vectors $a_1$ and $a_2$ as well as of the superlattice orientation angle $\varphi$.

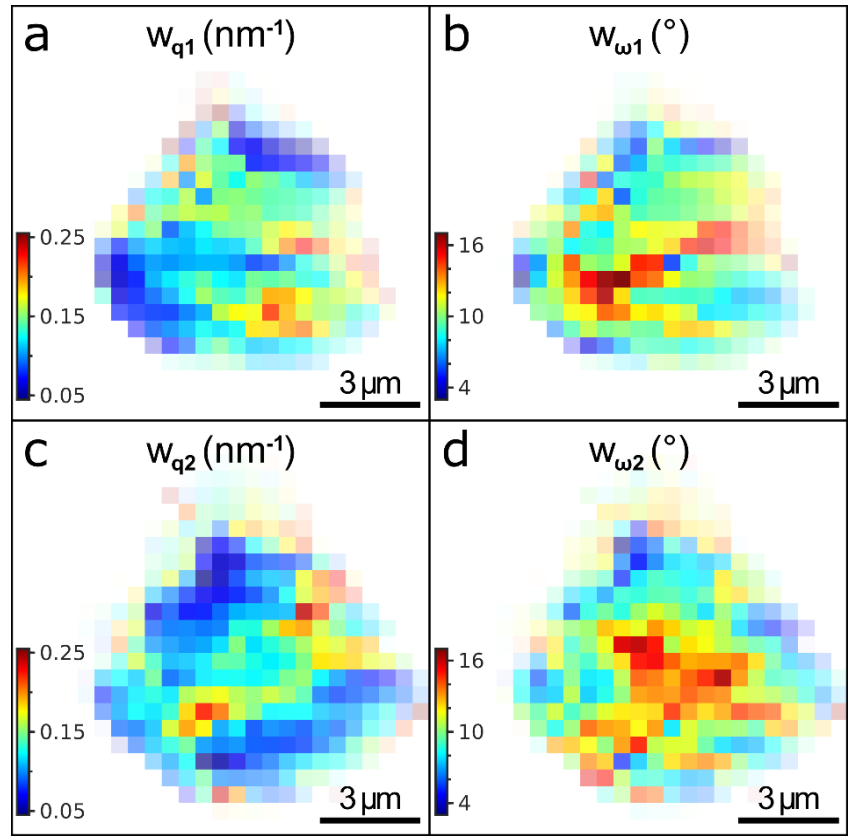

**Figure S13.** Mean extracted FWHMs in (a, c) radial and (b, d) azimuthal directions of (a,b) the 1st and 3rd and (c,d) the 2nd and 4th Bragg peaks. The pixel size is 500 nm.

The calculated lengths of the lattice vectors $a_1$ and $a_2$ are shown in **Fig. S14**. The distance between the adjacent NCs decrease in both directions on the edges of the supercrystal, but the effect is higher in the direction parallel to the nearest supercrystal edge. For example, in the point 1 in **Fig. S14**, the distance along $a_1$, which is pointing to the top-right parallel to the nearest edge, is smaller than along $a_2$, which is normal to the nearest edge. On the contrary, in the point 2, where $a_2$ is parallel to the nearest edge, the distance along this direction is smaller than along $a_1$. The anisotropy of the lattice shrinkage is better visible on the map of the $a_2/a_1$ ratio that is shown and discussed in the main text.

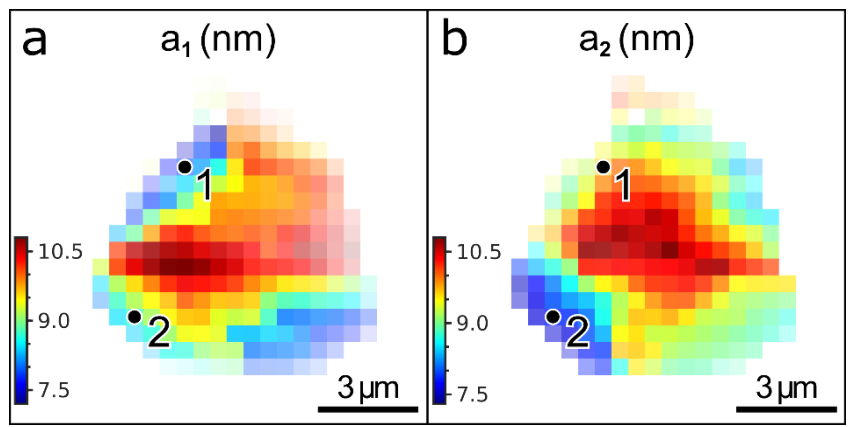

**Figure S14.** Maps of the calculated lengths of the lattice vectors $a_1$ and $a_2$. The points 1 and 2 are discussed in the text. The pixel size is 500 nm.

**Section S6. NCs orientation inside the SL**

In WAXS region, there are four Bragg peaks present on most of the diffraction pattern, as shown on the average one in **Fig. S15**. Analogous to the SAXS analysis, we fitted each of the peaks separately for each spatial point within the sample with the 2D Gaussian functions (**Eq. S3**).

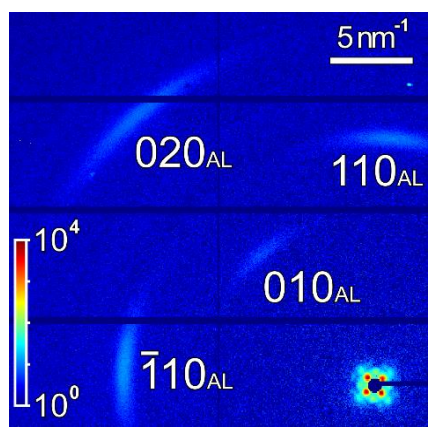

**Figure S15.** Average WAXS diffraction pattern with four prominent Bragg peaks. The indexes are given for pseudocubic atomic lattice oriented along [001] axis.

The intensities of the Bragg peaks extracted this way are shown in **Fig. S16**. $010_{AL}$, $020_{AL}$, $110_{AL}$ and $\bar{1}10_{AL}$ peaks are detected in the spatial points within the supercrystal, though lower intensity of the WAXS reflections did not allow detecting them on the very edges. The $010_{AL}$ reflection is registered on even smaller area, because it has the lowest intensity among the peaks.

The intensities are not uniform within the supercrystal that can indicate different thickness of the sample or slight out-of-plane rotation of the NCs. The intensities of $010_{AL}$, $020_{AL}$ and $110_{AL}$ reflections change the same way, while the intensity of $\bar{1}10_{AL}$ reflection stay almost constant. The inhomogeneous thickness would lead to the similar changes in intensities of all reflections, thus, the change most likely is due to the out-of-plane rotation of the NCs. The constant intensity of $\bar{1}10_{AL}$ reflection indicates that the rotation happens around an axis close to the $[\bar{1}10]_{AL}$ one. The changes in intensity of $010_{AL}$ and $020_{AL}$ reflections are qualitatively similar to the changes in intensity $I_2$ of $010_{SL}$ reflection (see **Fig. S12d**) that indicates simultaneous rotation of the NCs and SL keeping their mutual orientation. The changes in SAXS intensity are smaller due to lower effect of the Ewald's sphere curvature.

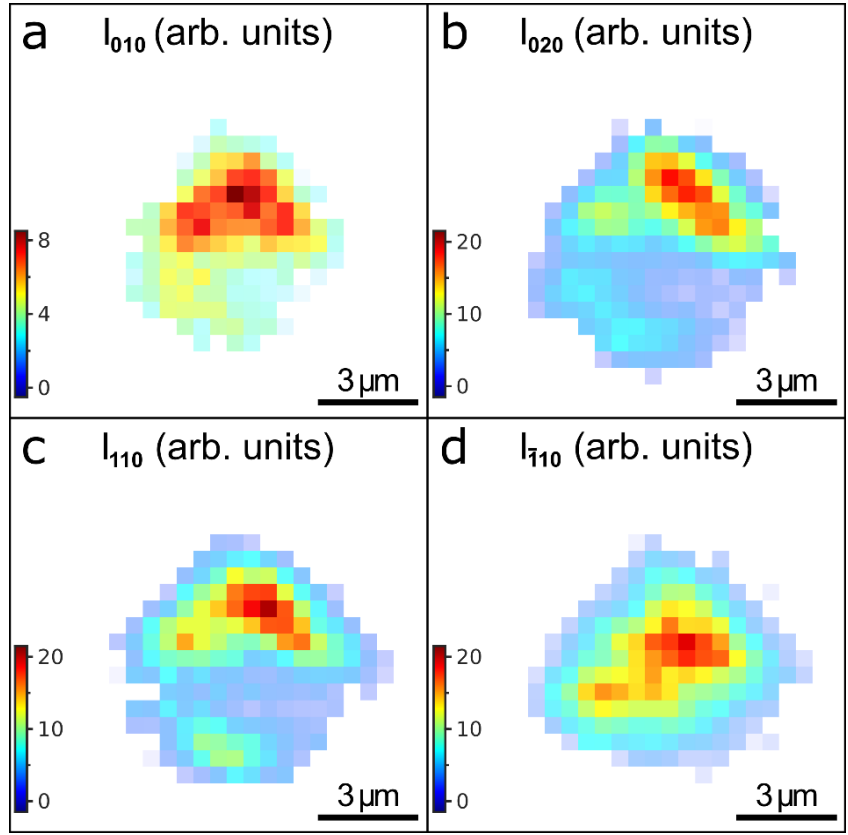

**Figure S16**. Extracted intensities of the WAXS Bragg peaks: **a**) $010_{AL}$, **b**) $020_{AL}$, **c**) $110_{AL}$ and **d**) $\bar{1}10_{AL}$. The pixel size is 500 nm.

The mean extracted momentum transfer values are $q_{010} = 10.87 \pm 0.05$ nm$^{-1}$, $q_{110} = 10.85 \pm 0.05$ nm$^{-1}$, $q_{\bar{1}10} = 10.91 \pm 0.03$ nm$^{-1}$, $q_{020} = 21.86 \pm 0.04$ nm$^{-1}$, that is in a good agreement with the values obtained from the average radial profiles discussed above in **Section S5**. The errorbars here represent the standard deviation of the values across the sample. The momentum transfer values do not depend on the spatial position on the sample, as shown in **Fig. S17**. The deviation is due to the noise together with the low intensities of the peaks themselves.

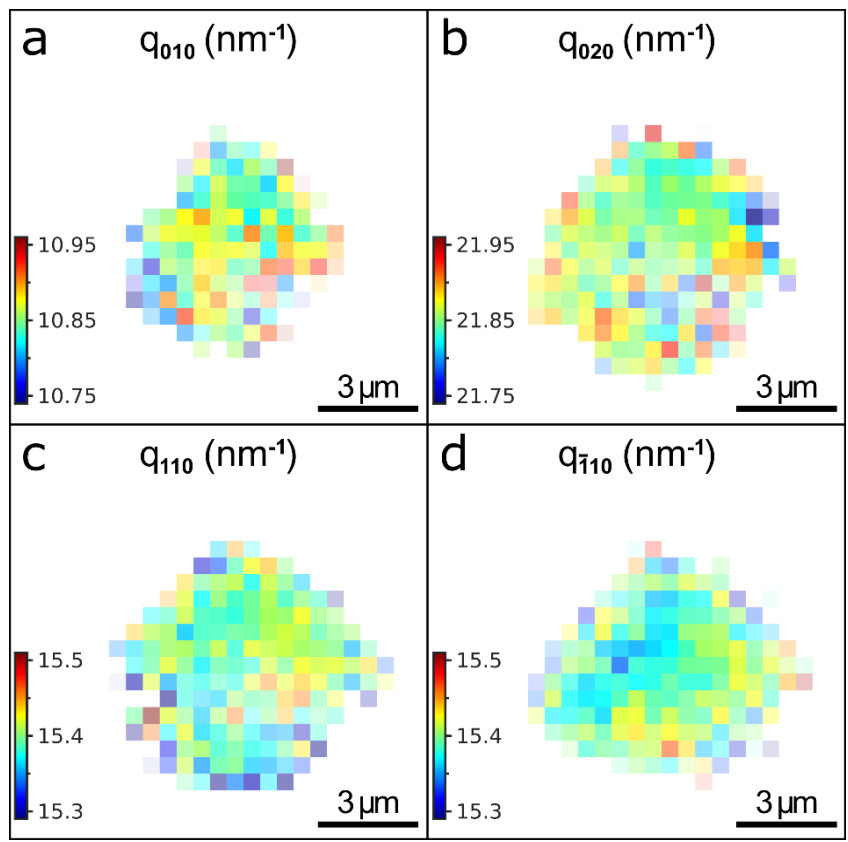

**Figure S17** Extracted momentum transfers of the WAXS Bragg peaks: **a**) $010_{AL}$, **b**) $020_{AL}$, **c**) $110_{AL}$ and **d**) $\bar{1}10_{AL}$. The pixel size is 500 nm.

The calculated from the q-values unit cell parameter is shown in **Fig. S18a**. It does not change within the supercrystal and remains constant at the value of $a_{AL} = 0.576\pm0.002$ nm. To better visualize the change in the unit cell parameter, we plot the values for each pixel against the distance of this pixel to the nearest supercrystal edge in **Fig. S18b**.

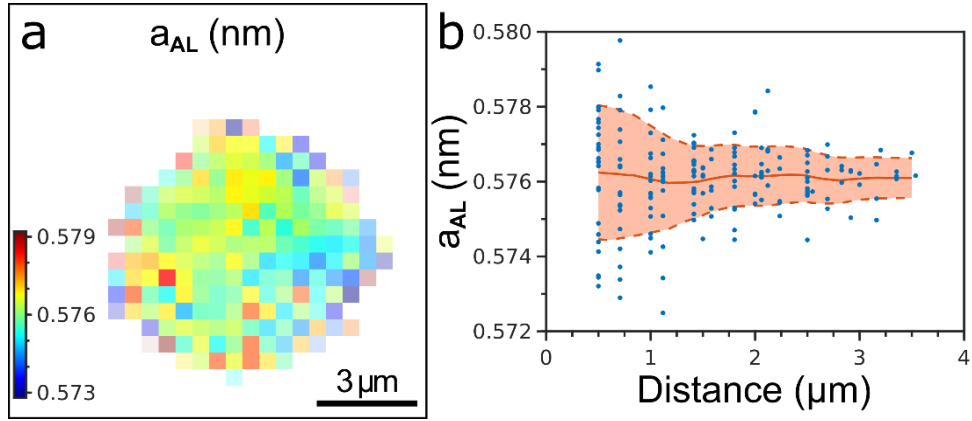

**Figure S18. a**) Calculated unit cell parameter $a_{AL}$ of the pseudo-cubic atomic lattice of the NCs and **b**) the same value for each pixel against the distance from this pixel to the nearest edge of the supercrystal. The red line shows the mean value, the dashed lines indicate the confidence interval of $\pm\sigma$. The pixel size in a) is 500 nm.

The FWHMs in radial direction, shown in **Fig. S19**, do not show any correlations with the spatial position on the sample as well. The mean values are $w_{q,010} = 0.73\pm0.10$ nm$^{-1}$, $w_{q,110} = 0.75\pm0.09$ nm$^{-1}$, $w_{q,\bar{1}10} = 0.73\pm0.07$ nm$^{-1}$ and $w_{q,020} = 0.85\pm0.11$ nm$^{-1}$, that is in good agreement with the values from the average radial profile discussed above in **Section S4**. The errorbars here represent the standard deviation of the values across the sample.

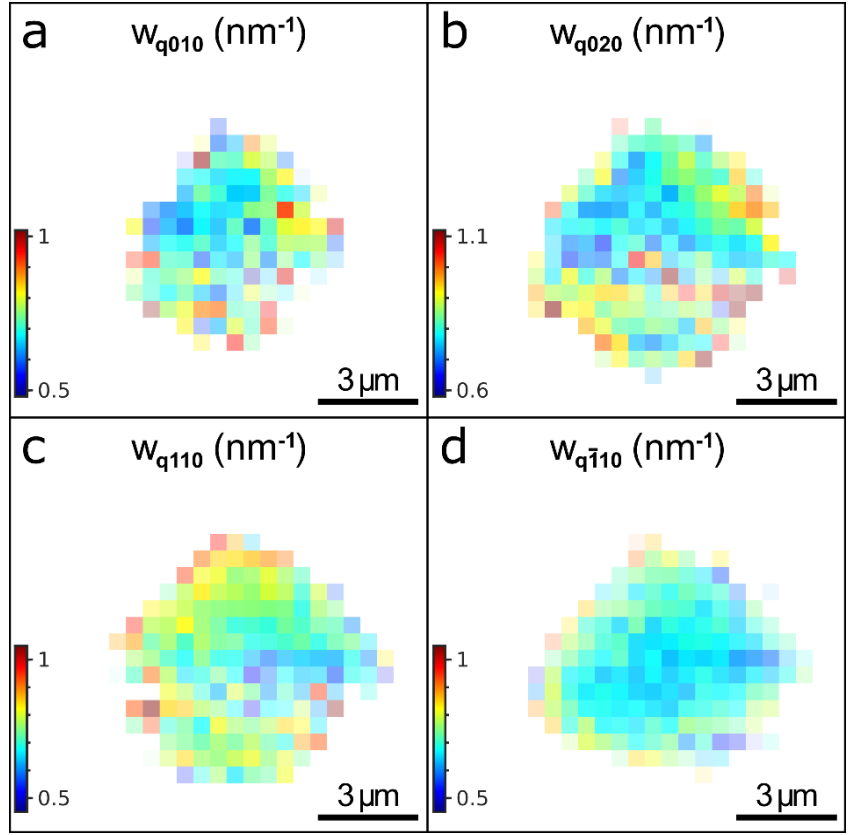

**Figure S19**. Extracted radial FWHMs of the WAXS Bragg peaks: **a**) $010_{AL}$, **b**) $020_{AL}$, **c**) $110_{AL}$ and **d**) $\bar{1}10_{AL}$. The pixel size is 500 nm.

We used the radial FWHMs to extract the atomic lattice distortion of the NCs by the Williamson-Hall method (**Eq. S2**). The lattice distortion $g_q$ (the ratio $\delta a_{AL}/a_{AL}$ of the FWHM $\delta a_{AL}$ of the unit cell parameter distribution around the mean value $a_{AL}$) was calculated as follows:

$$g_q = \langle \frac{1}{q_i}\sqrt{(w_i)^2 - \left(\frac{2\pi K}{L}\right)^2} \rangle_i, \qquad \text{(Eq. S10)}$$

where $w_i$ are the radial FWHMs for the present peaks, $q_i$ are the momentum transfer values of the corresponding peaks, $K$ is the shape constant, $L$ is the NCs size and the averaging is performed over all present Bragg peaks. The shape constant $K$ was discussed in **Section S4**; the NCs size was fixed at $L = 6.8$ nm obtained from the radial profile as described in the same **Section S4**. The resulting values of the atomic lattice distortion are shown in **Figure 6** in the main text. The mean value is $g_q = 1.5 \pm 0.9\%$ that is in good agreement with the values obtained from the average radial profile discussed in Section S1. The distortion gets slightly higher on the edges of the supercrystal that can be explained by the contraction of the NCs together with the superlattice. To better visualize the change in distortion, we plot the values for each pixel against the distance of this pixel to the nearest supercrystal edge in **Fig. 6b** in the main text. The atomic lattice distortion grows from < 1% in the middle of the supercrystal up to > 2% on the edges. The trend is even more evident for another sample, described in **Section S7**. Here the effect is less pronounced, probably, because of lower intensity of the WAXS Bragg peaks causing higher noise.

Since the NCs have pseudocubic atomic lattice (all angles are equal to 90°), the azimuthal position of the WAXS Bragg peaks directly correspond to the azimuthal orientation of the supercrystal unit cell basis vectors **a₁** and **a₂.** Thus, the azimuthal positions in real space are equal to the azimuthal positions in reciprocal space $\varphi_{hkl} = \omega_{hkl}$. The extracted azimuthal positions for all four peaks are shown in **Fig. S20**. The NCs are clearly rotating in-plane as soon as the positions change for different peaks together from point to point. As expected, the difference between the positions remains constant, e.g. $\varphi_{\bar{1}10} - \varphi_{110} = 90°$, $\varphi_{\bar{1}10} - \varphi_{020} = 45°$ etc. We calculated the average azimuthal NC position $\psi$ collinear to the $010_{AL}$ vector as $\psi = \langle \varphi_{010}, \varphi_{020}, \varphi_{110} + 45°, \varphi_{\bar{1}10} - 45° \rangle$, where the angle brackets denote averaging over the four angles. The angle ψ is used to study the azimuthal position of the NCs. It is shown in **Fig. 5c** and discussed in the main text.

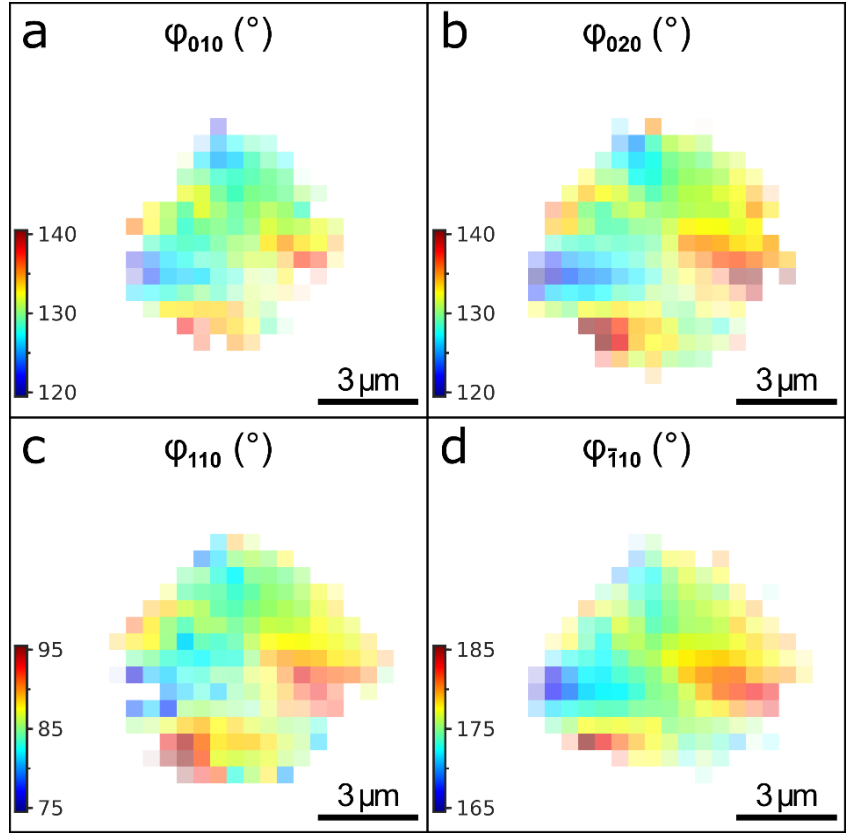

**Figure S20**. Extracted azimuthal positions of the WAXS Bragg peaks: **a**) $010_{AL}$, **b**) $020_{AL}$, **c**) $110_{AL}$ and **d**) $\bar{1}10_{AL}$. The positions are counted counterclockwise from a horizontal axis pointing to the right. The pixel size is 500 nm.

We studied the relative azimuthal orientation of the NCs and the SL comparing the azimuthal positions of the lattice vectors obtained from the Bragg peak analysis described above. The differences in angles between the $[010]_{AL}$ axis of the NCs and the mean line $M$ between the superlattice vectors $a_1$ and $a_2$ $\varDelta = \psi - \varphi$; between the $[010]_{AL}$ and $a_1$ $\varDelta_1 = \psi - \varphi_1$; between the $[010]_{AL}$ and $a_2$ $\varDelta_2 = \psi - \varphi_2$ are shown in **Fig. S21**. Clearly, the angle $\varDelta$ has narrower distribution with the mean value $\langle\varDelta\rangle = 45.2\pm1.7°$. For comparison, the angles $\varDelta_1$ and $\varDelta_2$ have broader distribution with the mean values $\langle\varDelta_1\rangle = 90.1\pm3.5°$ and $\langle\varDelta_2\rangle = 0.8\pm3.8°$. This is rather expected, because the angles between the NC crystallographic axes are constant, while the angle between the superlattice vectors changes from point to point in relatively broad range, as shown in **Fig. 4d** and discussed in the main text. At the same time, the angle between the mean lines between the superlattice vectors (e. g. *M* between *a₁* and *a₂* and *M'* between *a₁* and -*a₂*) is always equal to 90° and does not depend on the length and orientation of the vectors *a₁* and *a₂*. It makes possible keeping the mutual orientation between the $[100]_{AL}$ and $[010]_{AL}$ axes and the mean lines *M* and *M'*.

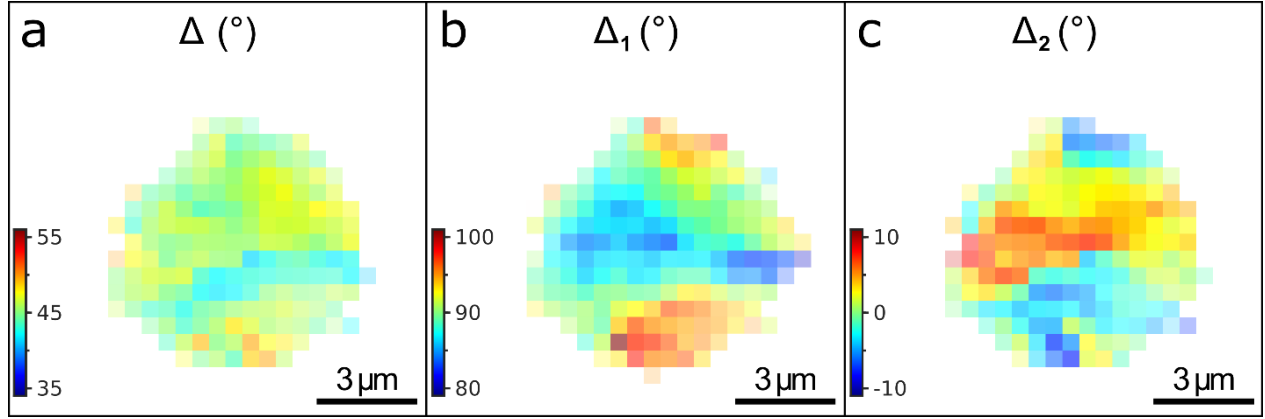

**Figure S21**. Relative angle between the direct lattice vectors of the NCs and the SL: **a**) between $[010]_{AL}$ and the mean line *M* between the *a₁* and *a₂*; **b**) between $[010]_{AL}$ and *a₁*; **c**) between $[010]_{AL}$ and *a₂*. The pixel size is 500 nm.

The azimuthal FWHMs shown in **Fig. S22** show clear dependence on the spatial position within the supercrystal. The FWHMs grow from 12° in the middle of the supercrystal up to 24° on the edges indicating higher azimuthal disorder of the NCs there.

We used the azimuthal FWHMs to extract the NCs angular disorder by the Williamson-Hall method. In this case, the lattice distortion *g* from **Eq. S2** is the FWHM $\delta\psi$ of angular distribution of the NCs around their mean position given by the azimuthal peak positions. The FWHM $\delta\psi$ was calculated as follows:

$$\delta\psi = \langle \frac{1}{q_i}\sqrt{(w_i q_i)^2 - \left(\frac{2\pi K}{L}\right)^2} \rangle_i,  \qquad \text{(Eq. S11)}$$

where $w_i$ are the azimuthal FWHMs (in radians) for the present peaks, $q_i$ are the momentum transfer values of the corresponding peaks, *K* is the shape constant, *L* is the NCs size and the averaging is performed over all present Bragg peaks. The shape constant *K* was discussed in **Section S4**; the NCs size was fixed at *L* = 6.8 nm obtained from the radial profile as described in the same **Section S4**. The resulting values of the FWHM are shown in **Figure 5d** and discussed in the main text.

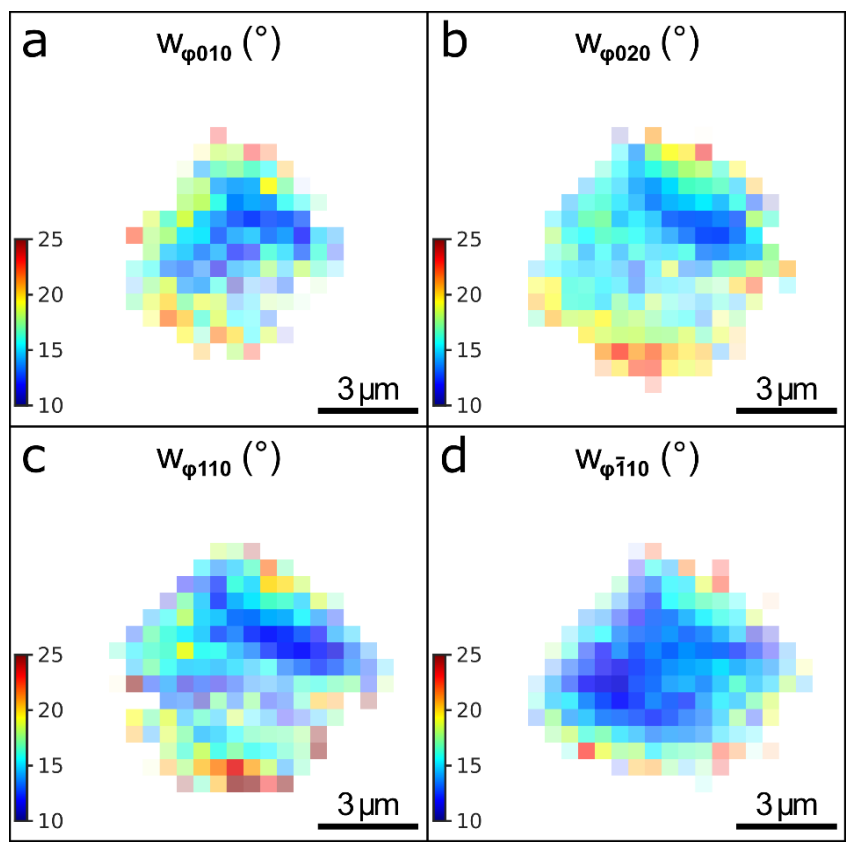

**Figure S22**. Extracted azimuthal FWHMs of the WAXS Bragg peaks: **a**) $010_{AL}$, **b**) $020_{AL}$, **c**) $110_{AL}$ and **d**) $\bar{1}10_{AL}$. The pixel size is 500 nm.

**Section S7. Another example of a supercrystal**

All studied supercrystals behave similarly. Here we present the main results obtained for one more supercrystal. The average WAXS and SAXS patterns as well as the SAXS-based diffraction map are shown in **Fig. S23**. The patterns look very similar to the ones observed for the sample described in the main text. The local structure was analyzed the same way as described above. The only difference is the WAXS analysis, performed only for three peaks $010_{AL}$, $020_{AL}$ and $\bar{1}10_{AL}$. The fourth $110_{AL}$ peak hardly fitted into the detector in this case and was excluded from the consideration.

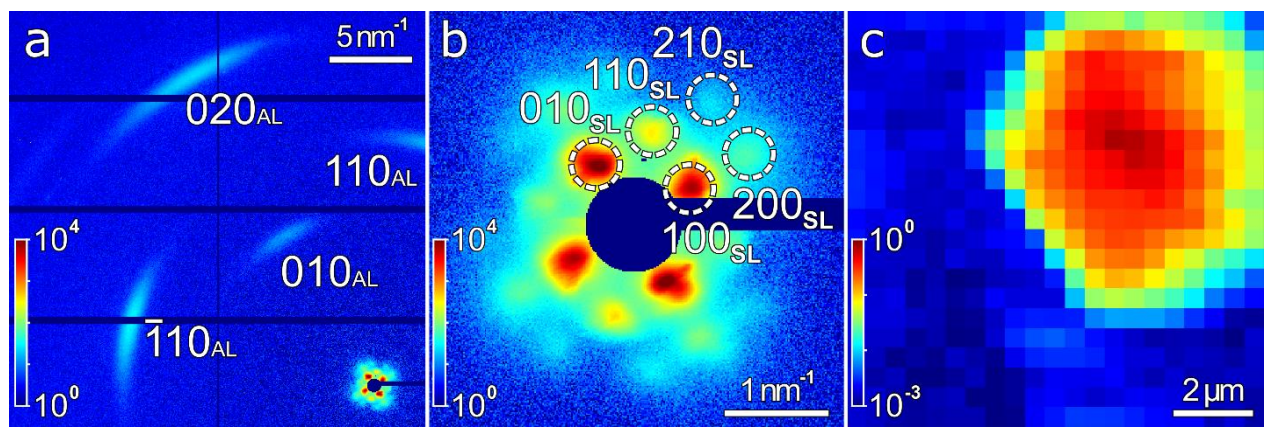

**Figure S23**. Average WAXS **a**) and SAXS **b**) patterns and **c**) SAXS-based diffraction map of the second sample. The Bragg peaks are indexed according to the "pseudocubic" structure of the atomic lattice and simple cubic structure of the superlattice. The pixel size in c) is 500 nm.

The main extracted parameters of the superlattice are shown in **Fig. S24**. The main observations are the same as for the sample described in the main text. The superlattice anisotropically shrinks on the edges of the supercrystal. The contraction happens preferentially in the directions parallel to the nearest supercrystal edge. The angle between the basis vectors $a_1$ and $a_2$ changes in the range of 75-105° and the superlattice rotates in-plane as indicated by the azimuthal position $\varphi$ of the mean line $M$ between $a_1$ and $a_2$.

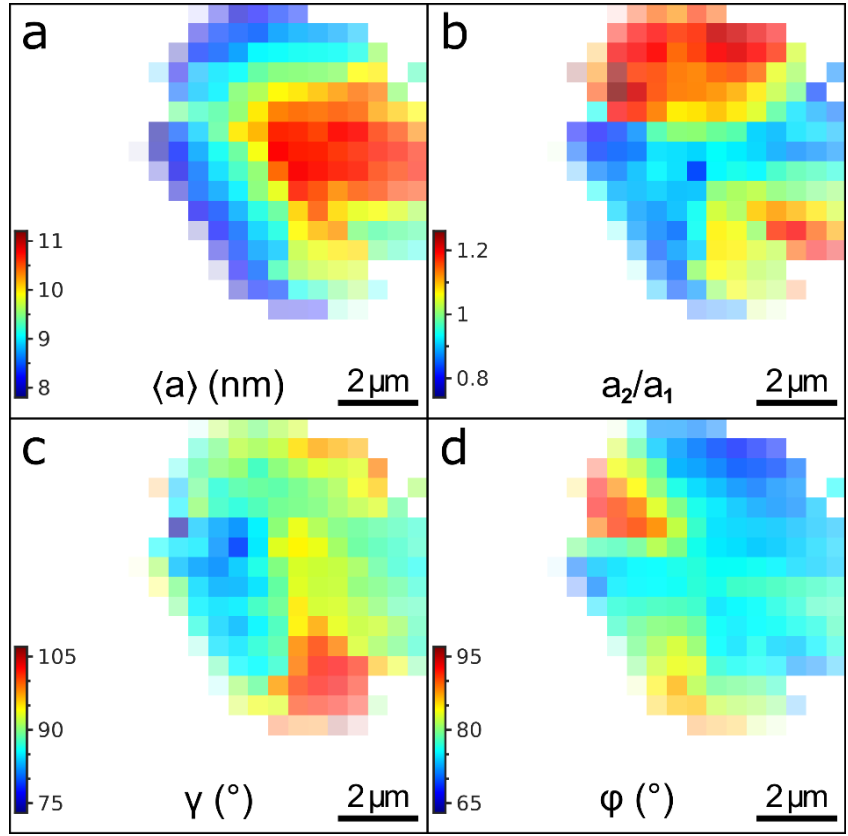

**Figure S24.** Extracted superlattice parameters: **a)** average unit cell parameter $\langle a \rangle = (a_1+a_2)/2$; **b)** ratio $a_2/a_1$ of the NC spacings along the basis vectors $a_2$ and $a_1$; **c)** angle $\gamma$ between the basis vectors $a_1$ and $a_2$; **d)** azimuthal position $\varphi$ of the mean line $M$ between the basis vectors $a_1$ and $a_2$. The pixel size is 500 nm.

The main extracted parameters of the atomic lattice are shown in **Fig. S25**. The main difference with the sample described in the main text is the homogeneous intensity of the WAXS Bragg peaks decreasing on the edges. It indicates absence of the out-of-plane rotations of the NCs. But the NCs rotate in-plane correlated with the mean line $M$ between the $a_1$ and $a_2$ superlattice vectors as can be seen from the difference angle $\Delta = \psi - \varphi$. The angular disorder of the NCs grows on the edges the same way as for the sample described in the main text.

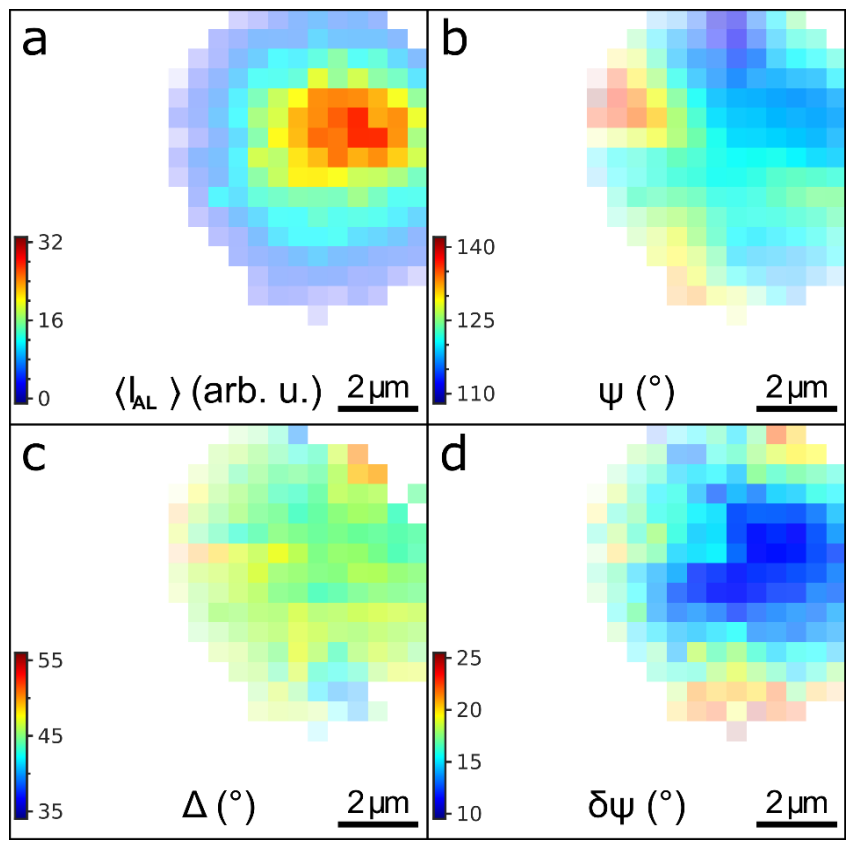

**Figure S25**. Extracted atomic lattice parameters: **a**) Average intensity of the WAXS Bragg peaks <$I_{AL}$>; **b**) azimuthal position $\psi$ of the 100$_{AL}$ crystallographic directions of the NCs; c) the relative angle $\Delta$ between the [010]$_{AL}$ axis and the mean line **M** between the $a_1$ and $a_2$ basic vectors of the SL; **d**) FWHM $\delta\psi$ of the angular disorder of the NCs around the mean azimuthal position $\psi$. The pixel size is 500 nm.

The calculated unit cell parameter of the atomic lattice from the *q*-values of the WAXS Bragg peaks is shown in **Fig. S26**. It doesn't show any correlation with the spatial position within the supercrystal and remains constant at the value of $a_{AL}$ = 0.576±0.002 nm.

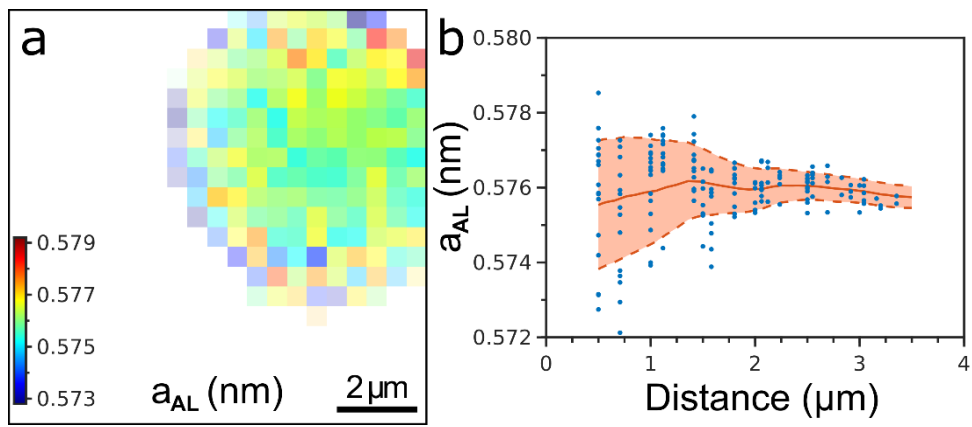

**Figure S26. a**) Calculated unit cell parameter $a_{AL}$ of the pseudo-cubic atomic lattice of the NCs and **b**) the same value for each pixel against the distance from this pixel to the nearest edge of the supercrystal. The red line shows the mean value, the dashed lines indicate the confidence interval of ±σ. The pixel size in a) is 500 nm.

The atomic lattice distortion was calculated from the FWHMs of the present WAXS Bragg peaks by the Williamson-Hall method (**Eq. S10**). The resulting values of the atomic lattice distortion are shown in **Figure S27**. The distortion gets slightly higher on the edges of the supercrystal that can be explained by the contraction of the NCs together with the superlattice. The atomic lattice distortion grows from about 0.5% in the middle of the supercrystal up to about 2.5% on the edges.

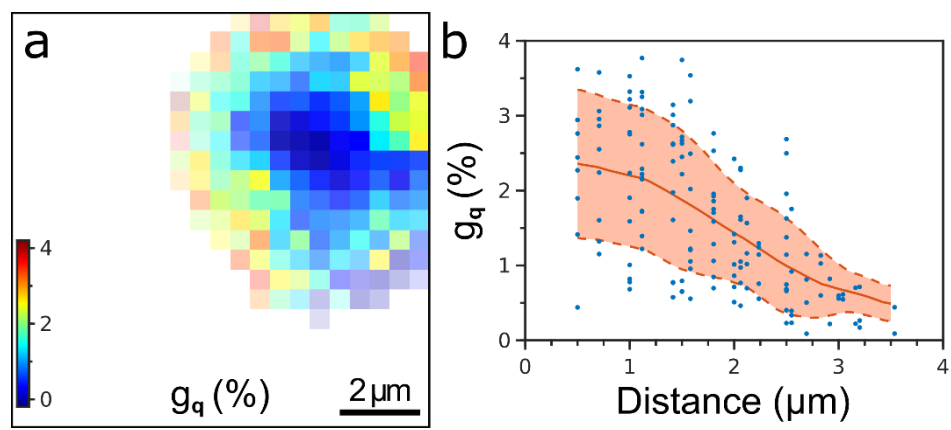

**Figure S27. a**) Atomic lattice distortion $g_q$ extracted from the radial FWHMs of the WAXS Bragg peaks by the Williamson-Hall method and **b**) the same value for each pixel against the distance from this pixel to the nearest edge of the supercrystal. The red line shows the mean value, the dashed lines indicate the confidence interval of ±σ. The pixel size in a) is 500 nm.

**Section S8. DFT-computed structures**

Computation protocol used in this work is described in the main text. Cartesian coordinates of all computed structures can be accessed from the coordinate file (.xyz).